\documentclass[%
 reprint,
 amsmath,amssymb,
 aps,
]{revtex4-2}

\usepackage{graphicx}
\usepackage{dcolumn}
\usepackage{bm}
\usepackage{color}

\begin{document}


\title{Nonclassical mechanical states in cavity optomechanics in the single-photon strong-coupling regime}

\author{Jonathan L. Wise}
\email{jonathan.wise@u-bordeaux.fr}
\author{Clément Dutreix}%
\author{Fabio Pistolesi}%

\affiliation{%
 Université de Bordeaux, CNRS, LOMA, UMR 5798, F-33400 Talence, France
}%




\date{\today}

\begin{abstract}


Generating nonclassical states of mechanical systems is a challenge relevant for testing the foundations of quantum mechanics and developing quantum technologies.
Significant effort has been made to search for such states in the stationary behaviour of cavity optomechanical systems.
We focus instead on the transient dynamics. 
We find that in the single-photon strong-coupling regime the presence of an optical drive causes an initial mechanical coherent state to evolve to a nonclassical state, with strongly negative Wigner function. 
An analytical treatment for weak drive reveals that these states are quantum superpositions of coherent states. 
Numerical simulation shows that the presence of Wigner negativity is robust against weak dissipation. 
%

\end{abstract}

\maketitle


\emph{Introduction. }
Ever since the early development of quantum mechanics and Schrödinger's provocative thought experiment \cite{schroedinger1935} physicists have been fascinated by the possibility of preparing macroscopic systems in nonclassical (NC) states. 
For this purpose, micro- and nanomechanical oscillators have emerged as promising candidates. 
Beyond a scientific curiosity, the generation and control of NC mechanical states will shed light on the foundations of quantum mechanics \cite{chen2013} and lead to the development of novel quantum technologies \cite{barzanjeh2022}, such as enhanced sensors \cite{qvarfort2018} and mechanical qubits \cite{pistolesi2021}. 
A very successful way to achieve precise control of a mechanical oscillator is by exploiting its ponderomotive coupling to a laser-driven optical cavity \cite{braginski1967, aspelmeyer2014}.
%
%
Significant advances include ground state cooling \cite{teufel2011, chan2011}, squeezing \cite{wollman2015, pirkkalainen2015, lecocq2015} and entanglement with either the cavity \cite{palomaki2013} or another oscillator \cite{ockeloen-korppi2018, riedinger2018, kotler2021}. 
These remarkable achievements have been obtained in the weak coupling limit, where the intrinsic nonlinear optomechanical (OM) coupling may be linearised. 
However, taking advantage of this intrinsic nonlinearity, one can expect to generate NC, \emph{non-Gaussian} states from semiclassical, \emph{Gaussian} input states. 
In cavity optomechanics, theoretical effort has been made to understand the single-photon strong-coupling regime, where the nonlinear coupling cannot be linearised  \cite{rabl2011, nunnenkamp2011, lemonde2013}.
%
%
Recent studies have predicted the formation of NC mechanical states -- quantum limit cycles for strong driving \cite{qian2012, nation2013, lorch2014} and cat states via application of a bichromatic laser \cite{hauer2023}.
While these studies focused on the steady state, surprisingly little attention has been paid to the transient dynamics.
%
Early works on the undriven cavity OM system showed that at discrete times the cavity field may evolve from an initial coherent state into a multi-component Schrödinger cat state \cite{mancini1997, bose1997, yurke1986}.
%
It was predicted that this optical nonclassicality may be projected onto the mechanical state.
However, this projection involves additional steps such as conditional measurements \cite{bose1997} and careful quantum superposition state preparation \cite{bose1999, marshall2003, pepper2012, hong2013, akram2013} of the cavity. 
%
\begin{figure}
\centering
\includegraphics[width = \columnwidth]{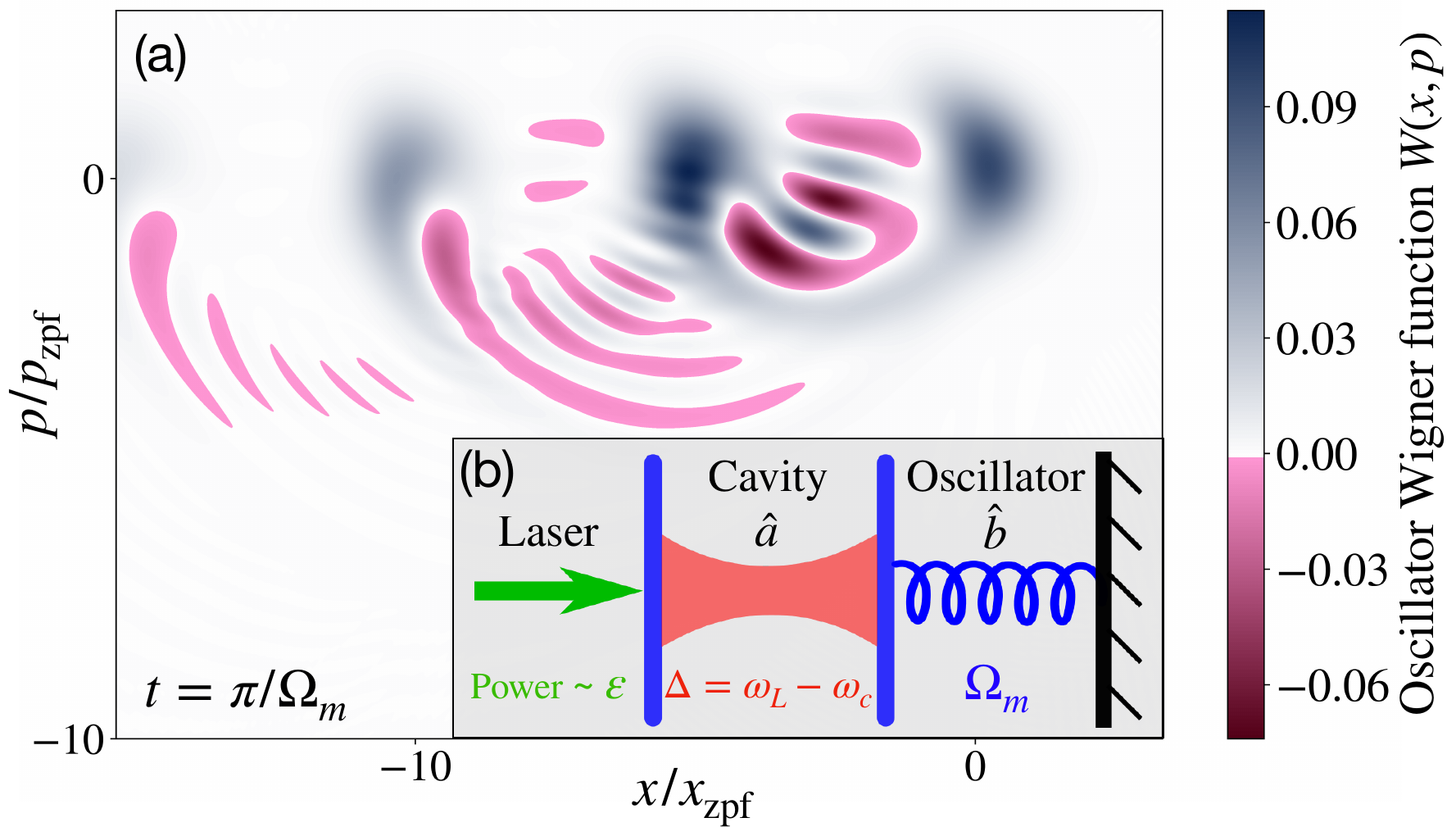}
\caption{(a) Wigner function of the mechanical oscillator after preparing an OM system in coherent states $|\alpha=1\rangle_a\otimes|\beta=0\rangle_b$ and driving the cavity for half a mechanical period [parameters given in Fig.~\ref{fig:dynamics_schematic}(d)]. (b) Schematic of the typical optically driven OM system considered.}
\label{fig:schematic_OM_setup}
\end{figure}

%
Here we show that NC mechanical states may be generated simply by weakly driving the optical cavity, circumventing the need for careful cavity state preparation, conditional measurements, strong driving or additional lasers. 
We propose the following steps: (i) Prepare the cavity and the oscillator in coherent states, and (ii) apply a weak laser drive to the cavity for a finite time, leading to the formation of a highly NC mechanical state [Fig.~\ref{fig:schematic_OM_setup}(a)]. 
Finally (iii) switch off the laser resulting in a periodic dynamics where the NC state reemerges every mechanical period. 
In the remainder of this Letter, we explain the rationale behind these steps. 
%
%
We present an analytical theory that accurately describes the NC states and makes explicit the physical mechanisms behind their development. 
%
Finally, we explore the dependence of the nonclassicality of the states as a function of the system parameters. 

\emph{Hamiltonian description. }
Since we study transient dynamics we begin by considering the unitary evolution.
The behaviour of a standard cavity OM system [Fig.~\ref{fig:schematic_OM_setup}(b)] is described (in the frame rotating at the laser drive frequency) by the Hamiltonian~\cite{mancini1994, law1995} 
\begin{equation}
\hat{H} = -\Delta\hat{a}^\dagger\hat{a} + \Omega_m\hat{b}^\dagger\hat{b} + g_0\hat{a}^\dagger\hat{a}\left(\hat{b}+\hat{b}^\dagger\right) + \varepsilon \left( \hat{a}+ \hat{a}^\dagger \right),
\label{eq:H}
\end{equation}
where $\hat{a}$ and $\hat{b}$ are annihilation operators for photons and phonons, respectively, and we set $\hbar=1$. 
We denote $\Delta=\omega_L-\omega_c$ the detuning between the laser frequency $\omega_L$ and the bare cavity frequency $\omega_c$, $\Omega_m$ the resonance frequency of the mechanical mode, $g_0$ the single-photon OM coupling strength, and $\varepsilon$ the laser drive strength. 
%
Applying the unitary transformation $\hat{U}_\mathrm{LF} = e^{\tilde{g}_0 \hat{a}^\dagger\hat{a}(\hat{b}^\dagger-\hat{b})}$ (known as Lang Firsov \cite{lang1963} or polaron \cite{nunnenkamp2011, mahan2000}), the Hamiltonian becomes
%
%
\begin{subequations}\label{eq:H_LF}
\begin{align}
\hat{H}_\mathrm{LF} &= \hat{H}_0 + \hat{V}, \tag{\ref{eq:H_LF}} \\ 
\hat{H}_0 &= -\Delta \hat{a}^\dagger\hat{a} - \mathcal{K}\left( \hat{a}^\dagger\hat{a}\right)^2 +\Omega_m\hat{b}^\dagger\hat{b}, \label{eq:H_0} \\
\hat{V} &= \varepsilon \left[\hat{D}(\tilde{g}_0)\hat{a}^\dagger + \hat{D}(-\tilde{g}_0)\hat{a}\right], \label{eq:V}
\end{align}
\end{subequations}
where $\tilde{g}_0= g_0/\Omega_m$, $\mathcal{K} = g_0^2/\Omega_m$ is the Kerr nonlinearity strength, and we introduced the mechanical displacement operator: $\hat{D}(\beta) = e^{\beta\hat{b}^\dagger - \beta^*\hat{b}}$. 
%
%
%

\emph{State initialisation. }
We choose as the initial state a coherent state in the laboratory frame for both light and mechanics: $\vert\psi(t=0)\rangle = \vert \alpha \rangle_a \otimes \vert \beta \rangle_b$. 
In a sideband resolved OM system the most natural way to achieve this is to employ radiation pressure cooling of the mechanical oscillator \cite{marquardt2007}. 
This prepares the cavity in a coherent state as desired and the oscillator in its quantum ground state ($\beta=0$), which could then be displaced by direct driving to achieve $\beta\ne 0$. 
We verify this approach by computing the stationary state of the driven-dissipative system (numerical methods described later), finding an average phonon number $\langle \hat{b}^\dagger\hat{b}\rangle = 0.01$, taking for example the parameters of Fig.~\ref{fig:extra_plot}(b) (but with $\varepsilon/\Omega_m = 0.2$ and $\Delta=-\Omega_m$).
%
%
While our results are general, for the plots in this Letter we take the initial state $\alpha = 1$ and $\beta = 0$ everywhere except in Fig.~\ref{fig:NC}(a). 
%

\begin{figure}
\centering
\includegraphics[width = \columnwidth]{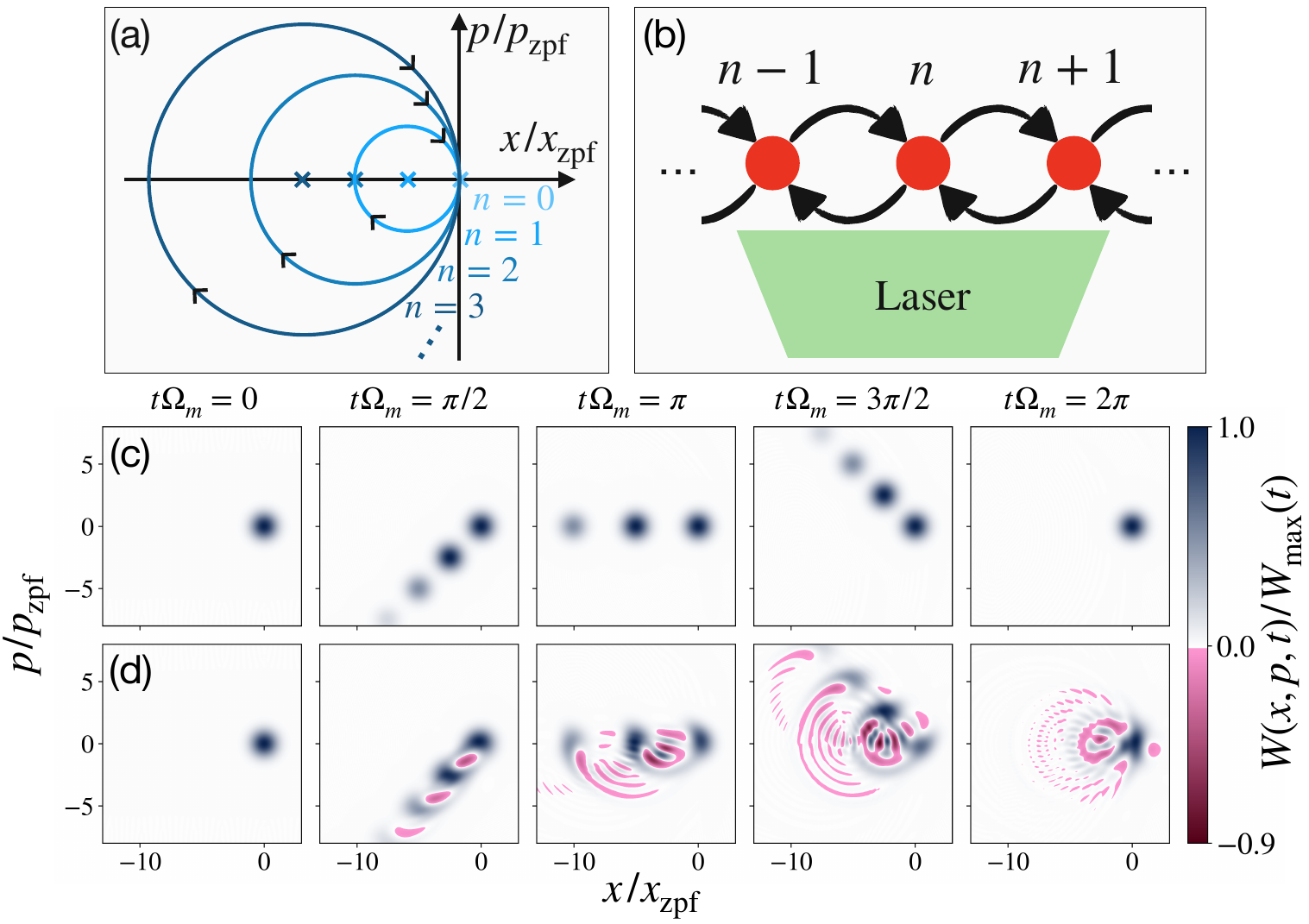}
\caption{Schematic visualisation of (a) the oscillator trajectories for the undriven system and (b) the cavity Fock state circulation induced by the laser drive. The oscillator Wigner function over one mechanical period for $g_0/\Omega_m = 1.8$ and $\Delta = 0$ in the undriven (c) and driven (d) case ($\varepsilon/\Omega_m = 0.3$), rescaled with respect to its maximum value $W_\mathrm{max}(t)$ -- see SM for absolute values. Note that $x_\mathrm{zpf}$ and $p_\mathrm{zpf}$ are scaled by a factor $\sqrt{2}$ with respect to their standard definition.}
\label{fig:dynamics_schematic} 
\end{figure}

\begin{figure}[b]
\centering
\includegraphics[width = \columnwidth]{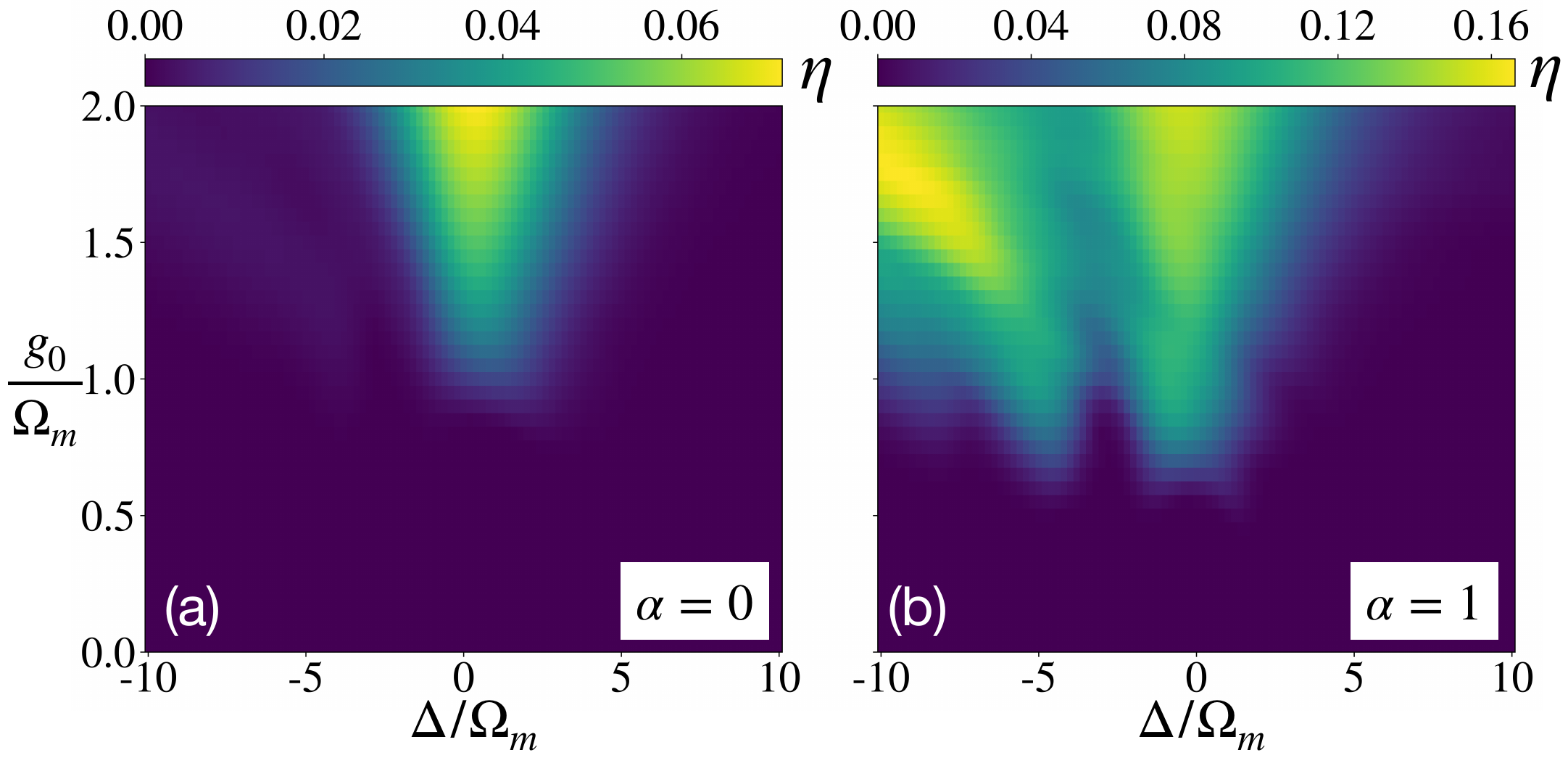}
\caption{Mechanical state nonclassical ratio $\eta$ as a function of system parameters after driving with $\varepsilon/\Omega_m=0.3$ for half a mechanical period choosing two different initial states: (a) $\vert \alpha =0\rangle_a \otimes\vert \beta =0\rangle_b$ and (b) $\vert \alpha =1\rangle_a \otimes\vert \beta =0\rangle_b$.}
\label{fig:NC} 
\end{figure}

\emph{Dynamics without optical drive. } 
It is instructive to start with the undriven system. 
The Hamiltonian (\ref{eq:H_0}) is diagonal, and so the time evolution can be obtained exactly.
In the original frame (\ref{eq:H}) the wavefunction reads
\begin{equation}
\vert \psi(t) \rangle = \sum_{n=0}^\infty a_n e^{i\phi_n(t)} \vert n \rangle_a \otimes \vert \beta_n(t) \rangle_b,
\label{eq:psi0}
\end{equation}
where $a_n = e^{-|\alpha|^2/2} \alpha^n/\sqrt{n!}$ specifies the initial cavity coherent state, $\beta_n(t) = e^{-i\Omega_m t}(\beta  + n \tilde{g}_0) -n \tilde{g}_0$ is the mechanical coherent state parameter, and the phase $\phi_n(t)$ is given in the Supplemental Material (SM). 
Analysing the state Eq.~(\ref{eq:psi0}) we see that each cavity Fock state $\vert n\rangle_a$ has become entangled with a distinct mechanical coherent state $\vert\beta_n(t)\rangle_b$. 
The latter describe oscillation trajectories with centres $x/x_\mathrm{zpf}=-n\tilde{g}_0$ and amplitudes $|\beta + n\tilde{g}_0|$, indicated pictorially in Fig.~\ref{fig:dynamics_schematic}(a). 
%
%
%
%
%
%
%
Tracing the system density matrix $\rho(t) = \vert\psi \rangle \langle \psi\vert$ over the oscillator subsystem reveals at discrete times a multi-component Schrödinger cat state for the cavity field \cite{bose1997}. 
This is a general feature of nonlinear oscillator Hamiltonians like Eq.~(\ref{eq:H_0})~\cite{yurke1986}, a fact that has been exploited to observe the formation and revival of transient cat states in nonlinear Kerr optical cavities \cite{kirchmair2013}. 

A natural question arises about the nature of the mechanical state when the photons are traced out. 
Doing so gives the reduced density matrix for the phonons
\begin{equation}
\rho_b(t) = \sum_{n=0}^\infty \left|a_n\right|^2 \vert\beta_n(t)\rangle \langle \beta_n(t)\vert.
\label{eq:rhob_undriven}
\end{equation}
This describes a statistical mixture of coherent states, i.e., a \emph{semiclassical} state. 
The presence of a NC state for the cavity but not the oscillator can be understood in the Lang-Firsov frame, where $\hat{H}_0$ describes an anharmonic cavity field, but a harmonic mechanical oscillator.
%
%
Thus an initial Gaussian state of the mechanical oscillator is bound to remain Gaussian. 
The evolution of the Wigner distribution of the state described by Eq.~(\ref{eq:rhob_undriven}), as shown in Fig.~\ref{fig:dynamics_schematic}(c), is always positive.
%
One way to obtain a non-Gaussian state for the oscillator would be if each photon Fock state was entangled with not just a single coherent state [as in Eq.~(\ref{eq:psi0})], but with a coherent superposition of mechanical states. 
In the following we show that this may be achieved by adding a cavity drive, which introduces an anharmonicity also for the mechanical oscillator, as shown in $\hat{V}$ [Eq.~(\ref{eq:V})].
%
%
%


\emph{Dynamics with weak optical drive. }
Let us now consider the full Hamiltonian of the driven system Eq.~(\ref{eq:H_LF}).
Figure~\ref{fig:dynamics_schematic}(d) shows the time evolution of the Wigner distribution, obtained by solving numerically the Schrödinger equation. 
All plots show results in the original basis of Eq.~(\ref{eq:H}).
The distribution clearly develops negative regions, a key signature of a NC non-Gaussian state.

We now provide some insight into this result. 
As illustrated in Fig.~\ref{fig:dynamics_schematic}(b), the drive couples the cavity Fock state $\vert n\rangle_a$ to the states $\vert n\pm 1\rangle_a$.
This should generate the desired coherent superposition of mechanical states entangled with a single photon Fock state. 
To show that, we calculate the time evolution of the wavefunction at first order in $\tilde{\varepsilon} = \varepsilon/\Omega_m$.
In the interaction picture in the Lang-Firsov basis defined in Eq.~(\ref{eq:H_LF}) the wavefunction obeys the Schrödinger equation $i\partial_t \vert\tilde{\psi}(t)\rangle = \hat{\tilde{V}}(t) \vert\tilde{\psi}(t)\rangle$,
where $\vert\tilde{\psi}(t)\rangle = e^{i\hat{H}_0t}\hat{U}_\mathrm{LF}\vert\psi(t)\rangle$ and $\hat{\tilde{V}}(t) = e^{i\hat{H}_0t}\hat{V}e^{-i\hat{H}_0t}$. 
%
This gives
\begin{align}
\vert\tilde{\psi}(t)\rangle \approx \mathcal{N}& \sum_{n=0}^\infty \vert n\rangle_a \otimes \left[A_n \vert\bar{\beta}_n \rangle_b \vphantom{\int_0^t} \right. \dots \nonumber \\
&\left. -i \tilde{\varepsilon} \int_0^{\Omega_m t} d\tilde{\tau}\, \left( \sqrt{n}A_{n-1}e^{i\varphi_{n-1}^+(\tau)} \vert\bar{\beta}_{n-1}^+(\tau)\rangle_b \right. \right. \nonumber \\ 
&\left.\vphantom{\int_0^t} \left.  + \sqrt{n+1}A_{n+1}e^{i\varphi_{n+1}^-(\tau)}  \vert\bar{\beta}_{n+1}^-(\tau)\rangle_b  \right) \right],
\label{eq:psi_LF}
\end{align}
with the prefactors $A_n = a_n e^{-i n\tilde{g}_0\mathrm{Im}\beta}$, the coherent state parameters $\bar{\beta}_n^\pm(\tau) = \bar{\beta}_n \pm \tilde{g}_0e^{i\Omega_m\tau}$ with $\bar{\beta}_n = \beta + n\tilde{g}_0$, and $\tilde{\tau} = \Omega_m \tau$.
The normalization factor $\mathcal{N}$ and the time-dependent phases $\varphi_n^\pm(\tau)$ are given explicitly in the SM. 
Equation~(\ref{eq:psi_LF}) indicates the formation of a NC mechanical state. 
%
%
Indeed, we see that each cavity state $\vert n\rangle_a$ is now entangled with \emph{multiple} mechanical states. 
Tracing the photons leads to a Wigner distribution which matches very well the numerical result shown in Fig.~\ref{fig:dynamics_schematic}(d) -- see SM for a direct comparison.

%


%
%

To elucidate further the resulting state we compute a short time ($\Omega_m t \ll 1$) expansion of the wavefunction Eq.~(\ref{eq:psi_LF}). 
Up to second order in time the wavefunction may be written in the form
%
%
%
\begin{equation}
\frac{\vert\tilde{\psi}(t)\rangle}{\mathcal{N}} \approx \sum_{n=0}^\infty \vert n\rangle_a\otimes\left[B_n^0 \vert\bar{\beta}_n\rangle_b  + B_n^+ \vert\bar{\beta}_{n+1}\rangle_b + B_n^-\vert\bar{\beta}_{n-1}\rangle_b \right],
\label{eq:psi_short_SP}
\end{equation}
where the coefficients $B_n^0 \approx A_n$ and $B_n^\pm \sim \tilde{\varepsilon} (\Omega_m t)^2$ (see SM for expressions). 
%
%
Equation~(\ref{eq:psi_short_SP}) shows that each photon state becomes entangled with a superposition of coherent states, whose origin is the drive-induced coupling to $\vert n\pm 1\rangle_a$. 
The mechanical part of the wavefunction may no longer be described by a single coherent state, and has been written in the form of a multi-component cat state for each $\vert n\rangle_a$.
%
%
In the region of validity of Eq.~(\ref{eq:psi_short_SP}) the Wigner distribution remains positive. 
However, this allows to infer the development of the superposition states that become strongly NC at later times, as seen in Fig.~\ref{fig:dynamics_schematic}(d). 
%

%
%
The full expression for the wavefunction (\ref{eq:psi_LF}) simplifies substantially if the initial state is the ground state: $\vert \psi(t=0)\rangle = \vert \alpha = 0 \rangle_a\otimes\vert \beta =0 \rangle_b$.
Transforming back to the Schrödinger picture and original (\ref{eq:H}) basis gives the expression $\vert \psi(t)\rangle = \mathcal{N}\left[ \vert \psi_0 \rangle - \tilde{\varepsilon} \vert \psi_1 \rangle\right]$, where $\vert \psi_0 \rangle = \vert 0\rangle_a \otimes \vert 0\rangle_b$ and the correction term is
\begin{equation}
\vert \psi_1 \rangle = \vert 1 \rangle_a \otimes \sum_{m=0}^\infty f_m(t) \hat{D}(-\tilde{g}_0) \vert m \rangle_b.
\end{equation}
We defined $f_m(t) = (\tilde{g}_0^m/\sqrt{m!})e^{-\tilde{g}_0^2/2}\left[1-e^{-iE_{1m}t}\right]/\tilde{E}_{1m}$, where $E_{1m}$ is the energy level of the first excited cavity state in Lang-Firsov basis, $E_{nm} = -\Delta n - \mathcal{K}n^2 + \Omega_m m$, and $\tilde{E}_{nm} = E_{nm}/\Omega_m$. 
Constructing the system density matrix and tracing out the photons leads to the oscillator state being given by a weighted superposition of displaced Fock states. 
Conveniently, the Wigner function may be written in terms of such states \cite{moya-cessa1993} and we find $W = \mathcal{N}^2 \left[W_0 + \tilde{\varepsilon}^2 W_1 \right]$, where $W_0(\xi) = (2/\pi)e^{-2|\xi|^2}$, $\mathrm{Re}\,\xi = x/x_\mathrm{zpf}$, and $\mathrm{Im}\,\xi = p/p_\mathrm{zpf}$. 
The correction term is:
\begin{equation}
W_1(\xi) = \frac{2}{\pi} \sum_{k=0}^\infty (-1)^k \left| \sum_{m=0}^\infty f_m(t) \langle \xi, k | -\tilde{g}_0, m \rangle\right|^2
\label{eq:W_1}
\end{equation}
where $\vert \xi, k\rangle = \hat{D}(\xi)\vert k \rangle$ is the displaced Fock state, whose overlap may be written exactly in terms of generalized Laguerre polynomials \cite{deoliveira1990, wunsche1991}. 
%
%
%
Although the structure of the expression is rather complex, we may use Eq.~(\ref{eq:W_1}) to show explicitly that the full Wigner function $W$ may become negative (see SM), a key signature of nonclassicality.

\emph{Quantifying the nonclassicality. }
%
%
The states predicted via our driving method are highly NC -- the maximum negative and positive amplitudes of the Wigner function can be of the same order [Fig.~\ref{fig:dynamics_schematic}(d)]. 
This degree of negativity reaches that of state-of-the-art predictions in the steady state \cite{qian2012, nation2013, lorch2014, hauer2023}. 
We further quantify the degree of nonclassicality via the NC ratio
\begin{equation}
\eta = \int_< \!\!\!dx \,dp\, \left| W(x,p)\right| \bigg/ \int_> \!\!\! dx\,dp\, W(x,p),
\end{equation}
where the symbol $<$ ($>$) indicates the integration domain where $W(x,p)$ is negative (positive) \cite{nation2013}. 
The state shown in Fig.~\ref{fig:schematic_OM_setup}(a) has $\eta = 0.15$. 
For comparison, we note that the Fock state $\vert 1\rangle$ has $\eta=0.18$ and a normalised cat-like superposition state $\propto \left(\vert\beta_0\rangle + \vert\beta_1\rangle\right)$ for the same parameters has $\eta=0.22$. 

The dependence of $\eta$ on the bare laser detuning $\Delta$ and OM coupling strength $g_0$ is shown in Fig.~\ref{fig:NC} for two different initial states. 
%
%
Choosing as the initial state the vacuum ($\alpha =0$), we observe in Fig.~\ref{fig:NC}(a) a peak in the Wigner negativity for zero detuning. 
Analyzing the structure of the corresponding analytical expression (\ref{eq:W_1}), this may be understood as a resonance of the transition from the initially populated ground $E_{00}=0$ state to the $E_{1m}$ state, where the largest Franck-Condon factor is with $m=\tilde{g}_0^2$ (see SM). 
%
%
Similarly we expect the additional structure present in Fig.~3(b) to be due to the $\vert 1 \rangle _a\rightarrow \vert 2 \rangle _a$ transition. 
%
%
%
The Wigner negativity appears primarily for negative detuning, contrary to what has been predicted for the  stationary state where nonclassicality is expected mainly for blue detuning \cite{ludwig2008, lorch2014}.
This intriguing difference may be explained by the fact that the transient NC states investigated here are very different in character, and consequently take place for different parameters. 

%



%

\begin{figure}
\centering
\includegraphics[width = \columnwidth]{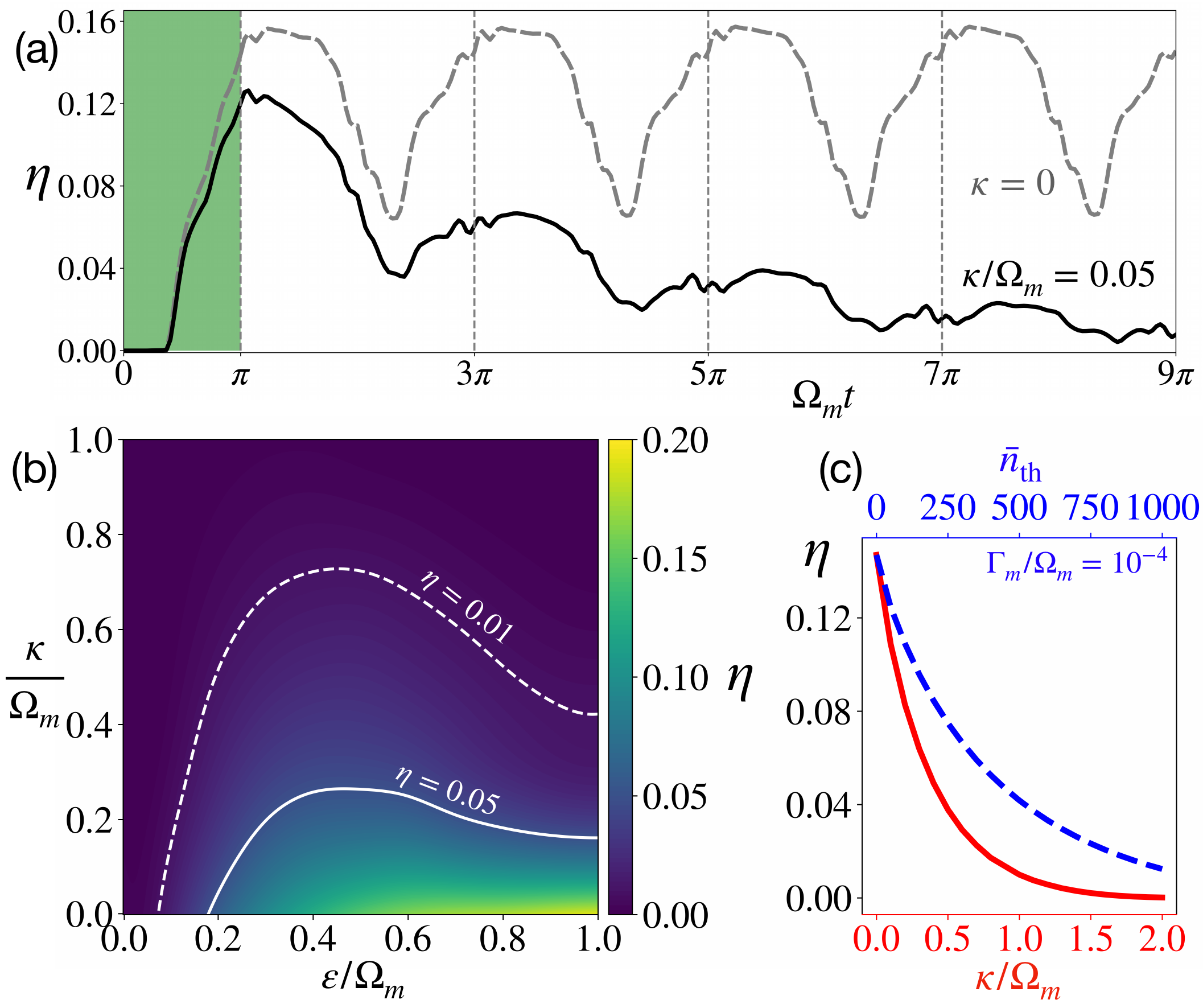}
\caption{(a) Evolution of $\eta$ during (shaded green, $\varepsilon/\Omega_m = 0.3$) and after ($\varepsilon = 0$) the drive pulse for zero dissipation (dashed grey) and for $\kappa/\Omega_m = 0.05$ and $\bar{n}_\mathrm{th}=10$, $\Gamma_m/\Omega_m = 10^{-4}$ (solid black). After driving for $t=\pi/\Omega_m$, we show (b) $\eta$ as a function of $\varepsilon$ and $\kappa$ (with $\Gamma_m/\Omega_m=10^{-4}$, $\bar{n}_\mathrm{th} = 0$), and (c) the effect of increasing cavity (solid red) and oscillator (dashed blue) dissipation rates, when the other is vanishing. We choose $\Delta=0$ in all the plots and $g_0/\Omega_m = 1$ in (b) and $g_0/\Omega_m = 1.8$ in (a) and (c). The plot (b) is an interpolation of numerical data -- see SM.}
\label{fig:NC_diss} 
\end{figure}

\begin{figure}
\centering
\includegraphics[width = \columnwidth]{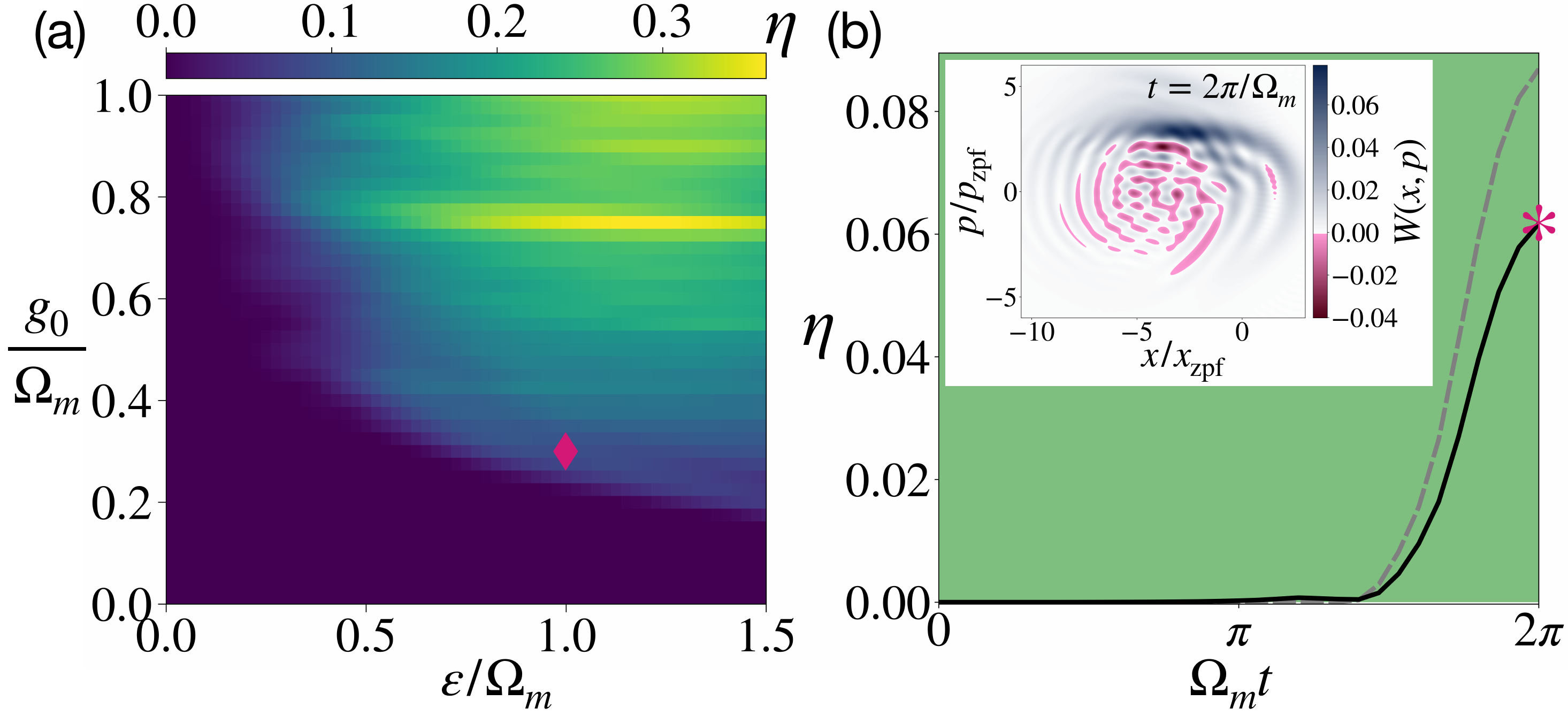}
\caption{(a) $\eta$ as a function of $g_0$ and $\varepsilon$ after driving for $t=2\pi/\Omega_m$ without dissipation. (b) Evolution of $\eta$ while driving for the parameters at $\blacklozenge$ ($g_0/\Omega_m=0.3$ and $\varepsilon/\Omega_m = 1$) for zero dissipation (dashed grey) and with $\kappa/\Omega_m=0.01$, $\bar{n}_\mathrm{th} = 10$, $\Gamma_m/\Omega_m=10^{-6}$ (solid black). Inset: Wigner function at $t=2\pi/\Omega_m$ at \scalebox{1.5}{$\ast$}.} 
\label{fig:extra_plot} 
\end{figure}

We note that if the drive is switched off, the subsequent evolution of the reduced density matrix of the mechanical oscillator is periodic in $2\pi/\Omega_m$ [see Eq.~(\ref{eq:rhob_undriven}) and SM]. 
This periodic revival of the NC state is reminiscent of the cat states produced in nonlinear cavities \cite{kirchmair2013}, and may in principle be leveraged when developing detection strategies.
The evolution of the NC ratio without dissipation is shown by the dashed grey line of Fig.~\ref{fig:NC_diss}(a).
The practical use of this periodicity depends on the effect of the dissipation. 
%
%

\emph{Dissipation. } 
%
We assess the effects of both optical and mechanical dissipation in the standard way, writing the OM master equation \cite{gardiner2004}
\begin{equation}
\frac{d\hat{\rho}}{dt} = -i[\hat{H}, \hat{\rho}] + \kappa \mathcal{D}[\hat{a}]\hat{\rho} + \Gamma_m(\bar{n}_\mathrm{th}+1)\mathcal{D}[\hat{b}]\hat{\rho} + \Gamma_m \bar{n}_\mathrm{th} \mathcal{D}[\hat{b}^\dagger]\hat{\rho}.
\label{eq:ME}
\end{equation}
Here, we assumed the cavity is coupled to a zero temperature optical bath with loss rate $\kappa$, while the mechanical resonator is in contact with an environment at temperature $T$ corresponding to mean occupation $\bar{n}_\mathrm{th} = 1/(e^{\hbar\Omega_m/k_BT}-1)$, with loss rate $\Gamma_m$.
The Lindblad dissipators are $\mathcal{D}[\hat{o}] = \hat{o} \hat{\rho} \hat{o}^\dagger - \left\{ \hat{o}^\dagger\hat{o}, \hat{\rho}\right\}/2$, with the curly braces denoting the anticommutator. 
We solve numerically \cite{johansson2012, johansson2013} the master equation (\ref{eq:ME}) for the full dissipative system dynamics. 
In Fig.~\ref{fig:NC_diss}(a) we illustrate the effect of dissipation on the full time evolution, when the drive is switched off after $t=\pi/\Omega_m$. 
In Fig.~\ref{fig:NC_diss}(b) we show the nonclassical ratio after driving for half a mechanical period, as a function of drive strength and cavity dissipation rate.
We observe that to a certain extent one may compensate for finite cavity dissipation by increasing the drive strength, and an optimum value $\varepsilon/\Omega_m \approx 0.5$ is predicted. 
As shown in Fig.~\ref{fig:NC_diss}(c) increasing the dissipation rate $\kappa$ or thermal bath temperature $T$ leads to a significant reduction of the nonclassicality achieved by driving. 

%
%

%




\emph{Experimental observability. }
So far we have shown results where $g_0 \gtrsim \Omega_m \gg \kappa, \Gamma_m$, known as the ``single-photon ultrastrong-coupling regime''.
However, we highlight in Fig.~\ref{fig:extra_plot} that this is not a necessary condition for the generation of nonclassical mechanical states via this method.
Indeed, using a drive pulse with stronger intensity and/or longer duration leads to highly nonclassical states even in the less experimentally stringent ``single-photon strong-coupling regime'', where $g_0 \gg \kappa, \Gamma_m$ but $g_0 < \Omega_m$.
This puts this method on par with other NC mechanical state generation methods \cite{qian2012, nation2013, lorch2014, hauer2023} in terms of the conditions on the optomechanical parameters.  
%
%
%

A crucial first step has been achieved by reaching the regime $g_0\gg \Omega_m$ with nanowire oscillators \cite{fogliano2021}, however for this system $\kappa\gg g_0$.
%
%
Other promising candidates for realising the strong and ultrastrong-coupling regimes are OM crystals \cite{chan2012, leijssen2015, bozkurt2022}, trampoline resonators \cite{reinhardt2016}, electromechanical systems \cite{teufel2011b, samanta2023}, and OM systems coupled to qubits \cite{heikkila2014, pirkkalainen2015a, manninen2022}.
%

%
The nature of a nonclassical state may be revealed experimentally by quantum state reconstruction. 
This can be done by performing quadrature measurements via tomography (see Ref.~\cite{vanner2015} for a review focusing on mechanical systems). 
%
For cavity optomechanical systems low noise mechanical quadrature measurements have been made via quantum nondemolition \cite{suh2014, lecocq2015} and transient amplification methods \cite{delaney2019}. 
Alternatively, dispersive coupling of a mechanical oscillator to a two-level system has allowed for measurement of the full Wigner function for Fock and superposition states via Ramsey interferometry \cite{chu2018, wollack2022, schrinski2023, bild2023}.
This is in direct analogy to what is done using atoms for state reconstruction of optical cavity fields \cite{haroche2006}. 
Quantum state reconstruction may also be achieved in the strong-coupling regime by measuring the cavity photon emission spectrum \cite{liao2014}. 
%
%
The periodic reemergence of our NC state once the drive is switched off may also allow one to read out the state via stroboscopic measurement. 



\emph{Conclusions. }
Cavity optomechanical systems have been investigated extensively for the generation of nonclassical mechanical states. 
Remarkably, we found that in the strong-coupling regime such a mechanical state may emerge spontaneously by simply sending a laser pulse to the cavity prepared in a coherent state. 
%
We showed that this transient state is a quantum superposition of coherent states, which will reemerge periodically after the drive pulse for weak dissipation.
%
%
%
%


\emph{Acknowledgments. }
We acknowledge support from the French Agence Nationale de la Recherche (Grant SINPHOCOM ANR-19-CE47-0012) and from the French government in the framework of the University of Bordeaux's France 2030 program / GPR LIGHT.



\bibliography{ms}

\begin{thebibliography}{63}%
\makeatletter
\providecommand \@ifxundefined [1]{%
 \@ifx{#1\undefined}
}%
\providecommand \@ifnum [1]{%
 \ifnum #1\expandafter \@firstoftwo
 \else \expandafter \@secondoftwo
 \fi
}%
\providecommand \@ifx [1]{%
 \ifx #1\expandafter \@firstoftwo
 \else \expandafter \@secondoftwo
 \fi
}%
\providecommand \natexlab [1]{#1}%
\providecommand \enquote  [1]{``#1''}%
\providecommand \bibnamefont  [1]{#1}%
\providecommand \bibfnamefont [1]{#1}%
\providecommand \citenamefont [1]{#1}%
\providecommand \href@noop [0]{\@secondoftwo}%
\providecommand \href [0]{\begingroup \@sanitize@url \@href}%
\providecommand \@href[1]{\@@startlink{#1}\@@href}%
\providecommand \@@href[1]{\endgroup#1\@@endlink}%
\providecommand \@sanitize@url [0]{\catcode `\\12\catcode `\$12\catcode
  `\&12\catcode `\#12\catcode `\^12\catcode `\_12\catcode `\%12\relax}%
\providecommand \@@startlink[1]{}%
\providecommand \@@endlink[0]{}%
\providecommand \url  [0]{\begingroup\@sanitize@url \@url }%
\providecommand \@url [1]{\endgroup\@href {#1}{\urlprefix }}%
\providecommand \urlprefix  [0]{URL }%
\providecommand \Eprint [0]{\href }%
\providecommand \doibase [0]{https://doi.org/}%
\providecommand \selectlanguage [0]{\@gobble}%
\providecommand \bibinfo  [0]{\@secondoftwo}%
\providecommand \bibfield  [0]{\@secondoftwo}%
\providecommand \translation [1]{[#1]}%
\providecommand \BibitemOpen [0]{}%
\providecommand \bibitemStop [0]{}%
\providecommand \bibitemNoStop [0]{.\EOS\space}%
\providecommand \EOS [0]{\spacefactor3000\relax}%
\providecommand \BibitemShut  [1]{\csname bibitem#1\endcsname}%
\let\auto@bib@innerbib\@empty
\bibitem [{\citenamefont {Schr{\"o}dinger}(1935)}]{schroedinger1935}%
  \BibitemOpen
  \bibfield  {author} {\bibinfo {author} {\bibfnamefont {E.}~\bibnamefont
  {Schr{\"o}dinger}},\ }\bibfield  {title} {\bibinfo {title} {{Die
  gegenw{\"a}rtige Situation in der Quantenmechanik (engl. transl. Trimmer JP
  1980 proc. Am. Phil. Soc. 124 323)}},\ }\href
  {https://doi.org/10.1007/BF01491891} {\bibfield  {journal} {\bibinfo
  {journal} {Naturwissenschaften}\ }\textbf {\bibinfo {volume} {23}},\ \bibinfo
  {pages} {807} (\bibinfo {year} {1935})}\BibitemShut {NoStop}%
\bibitem [{\citenamefont {Chen}(2013)}]{chen2013}%
  \BibitemOpen
  \bibfield  {author} {\bibinfo {author} {\bibfnamefont {Y.}~\bibnamefont
  {Chen}},\ }\bibfield  {title} {\bibinfo {title} {Macroscopic quantum
  mechanics: Theory and experimental concepts of optomechanics},\ }\href
  {https://doi.org/10.1088/0953-4075/46/10/104001} {\bibfield  {journal}
  {\bibinfo  {journal} {Journal of Physics B: Atomic, Molecular and Optical
  Physics}\ }\textbf {\bibinfo {volume} {46}},\ \bibinfo {pages} {104001}
  (\bibinfo {year} {2013})}\BibitemShut {NoStop}%
\bibitem [{\citenamefont {Barzanjeh}\ \emph {et~al.}(2022)\citenamefont
  {Barzanjeh}, \citenamefont {Xuereb}, \citenamefont {Gr{\"o}blacher},
  \citenamefont {Paternostro}, \citenamefont {Regal},\ and\ \citenamefont
  {Weig}}]{barzanjeh2022}%
  \BibitemOpen
  \bibfield  {author} {\bibinfo {author} {\bibfnamefont {S.}~\bibnamefont
  {Barzanjeh}}, \bibinfo {author} {\bibfnamefont {A.}~\bibnamefont {Xuereb}},
  \bibinfo {author} {\bibfnamefont {S.}~\bibnamefont {Gr{\"o}blacher}},
  \bibinfo {author} {\bibfnamefont {M.}~\bibnamefont {Paternostro}}, \bibinfo
  {author} {\bibfnamefont {C.~A.}\ \bibnamefont {Regal}},\ and\ \bibinfo
  {author} {\bibfnamefont {E.~M.}\ \bibnamefont {Weig}},\ }\bibfield  {title}
  {\bibinfo {title} {Optomechanics for quantum technologies},\ }\href
  {https://doi.org/10.1038/s41567-021-01402-0} {\bibfield  {journal} {\bibinfo
  {journal} {Nature Physics}\ }\textbf {\bibinfo {volume} {18}},\ \bibinfo
  {pages} {15} (\bibinfo {year} {2022})}\BibitemShut {NoStop}%
\bibitem [{\citenamefont {Qvarfort}\ \emph {et~al.}(2018)\citenamefont
  {Qvarfort}, \citenamefont {Serafini}, \citenamefont {Barker},\ and\
  \citenamefont {Bose}}]{qvarfort2018}%
  \BibitemOpen
  \bibfield  {author} {\bibinfo {author} {\bibfnamefont {S.}~\bibnamefont
  {Qvarfort}}, \bibinfo {author} {\bibfnamefont {A.}~\bibnamefont {Serafini}},
  \bibinfo {author} {\bibfnamefont {P.~F.}\ \bibnamefont {Barker}},\ and\
  \bibinfo {author} {\bibfnamefont {S.}~\bibnamefont {Bose}},\ }\bibfield
  {title} {\bibinfo {title} {Gravimetry through non-linear optomechanics},\
  }\href {https://doi.org/10.1038/s41467-018-06037-z} {\bibfield  {journal}
  {\bibinfo  {journal} {Nature Communications}\ }\textbf {\bibinfo {volume}
  {9}},\ \bibinfo {pages} {3690} (\bibinfo {year} {2018})}\BibitemShut
  {NoStop}%
\bibitem [{\citenamefont {Pistolesi}\ \emph {et~al.}(2021)\citenamefont
  {Pistolesi}, \citenamefont {Cleland},\ and\ \citenamefont
  {Bachtold}}]{pistolesi2021}%
  \BibitemOpen
  \bibfield  {author} {\bibinfo {author} {\bibfnamefont {F.}~\bibnamefont
  {Pistolesi}}, \bibinfo {author} {\bibfnamefont {A.~N.}\ \bibnamefont
  {Cleland}},\ and\ \bibinfo {author} {\bibfnamefont {A.}~\bibnamefont
  {Bachtold}},\ }\bibfield  {title} {\bibinfo {title} {Proposal for a
  {{Nanomechanical Qubit}}},\ }\href
  {https://doi.org/10.1103/PhysRevX.11.031027} {\bibfield  {journal} {\bibinfo
  {journal} {Physical Review X}\ }\textbf {\bibinfo {volume} {11}},\ \bibinfo
  {pages} {031027} (\bibinfo {year} {2021})}\BibitemShut {NoStop}%
\bibitem [{\citenamefont {Braginski}\ and\ \citenamefont
  {Manukin}(1967)}]{braginski1967}%
  \BibitemOpen
  \bibfield  {author} {\bibinfo {author} {\bibfnamefont {{\relax
  VB}.}~\bibnamefont {Braginski}}\ and\ \bibinfo {author} {\bibfnamefont
  {{\relax AB}.}~\bibnamefont {Manukin}},\ }\bibfield  {title} {\bibinfo
  {title} {Ponderomotive effects of electromagnetic radiation},\ }\href@noop {}
  {\bibfield  {journal} {\bibinfo  {journal} {Soviet Physics--JETP [translation
  of Zhurnal Eksperimentalnoi i Teoreticheskoi Fiziki]}\ }\textbf {\bibinfo
  {volume} {25}},\ \bibinfo {pages} {653} (\bibinfo {year} {1967})}\BibitemShut
  {NoStop}%
\bibitem [{\citenamefont {Aspelmeyer}\ \emph {et~al.}(2014)\citenamefont
  {Aspelmeyer}, \citenamefont {Kippenberg},\ and\ \citenamefont
  {Marquardt}}]{aspelmeyer2014}%
  \BibitemOpen
  \bibfield  {author} {\bibinfo {author} {\bibfnamefont {M.}~\bibnamefont
  {Aspelmeyer}}, \bibinfo {author} {\bibfnamefont {T.~J.}\ \bibnamefont
  {Kippenberg}},\ and\ \bibinfo {author} {\bibfnamefont {F.}~\bibnamefont
  {Marquardt}},\ }\bibfield  {title} {\bibinfo {title} {Cavity optomechanics},\
  }\href {https://doi.org/10.1103/RevModPhys.86.1391} {\bibfield  {journal}
  {\bibinfo  {journal} {Reviews of Modern Physics}\ }\textbf {\bibinfo {volume}
  {86}},\ \bibinfo {pages} {1391} (\bibinfo {year} {2014})}\BibitemShut
  {NoStop}%
\bibitem [{\citenamefont {Teufel}\ \emph
  {et~al.}(2011{\natexlab{a}})\citenamefont {Teufel}, \citenamefont {Donner},
  \citenamefont {Li}, \citenamefont {Harlow}, \citenamefont {Allman},
  \citenamefont {Cicak}, \citenamefont {Sirois}, \citenamefont {Whittaker},
  \citenamefont {Lehnert},\ and\ \citenamefont {Simmonds}}]{teufel2011}%
  \BibitemOpen
  \bibfield  {author} {\bibinfo {author} {\bibfnamefont {J.~D.}\ \bibnamefont
  {Teufel}}, \bibinfo {author} {\bibfnamefont {T.}~\bibnamefont {Donner}},
  \bibinfo {author} {\bibfnamefont {D.}~\bibnamefont {Li}}, \bibinfo {author}
  {\bibfnamefont {J.~W.}\ \bibnamefont {Harlow}}, \bibinfo {author}
  {\bibfnamefont {M.~S.}\ \bibnamefont {Allman}}, \bibinfo {author}
  {\bibfnamefont {K.}~\bibnamefont {Cicak}}, \bibinfo {author} {\bibfnamefont
  {A.~J.}\ \bibnamefont {Sirois}}, \bibinfo {author} {\bibfnamefont {J.~D.}\
  \bibnamefont {Whittaker}}, \bibinfo {author} {\bibfnamefont {K.~W.}\
  \bibnamefont {Lehnert}},\ and\ \bibinfo {author} {\bibfnamefont {R.~W.}\
  \bibnamefont {Simmonds}},\ }\bibfield  {title} {\bibinfo {title} {Sideband
  cooling of micromechanical motion to the quantum ground state},\ }\href
  {https://doi.org/10.1038/nature10261} {\bibfield  {journal} {\bibinfo
  {journal} {Nature}\ }\textbf {\bibinfo {volume} {475}},\ \bibinfo {pages}
  {359} (\bibinfo {year} {2011}{\natexlab{a}})}\BibitemShut {NoStop}%
\bibitem [{\citenamefont {Chan}\ \emph {et~al.}(2011)\citenamefont {Chan},
  \citenamefont {Alegre}, \citenamefont {{Safavi-Naeini}}, \citenamefont
  {Hill}, \citenamefont {Krause}, \citenamefont {Gr{\"o}blacher}, \citenamefont
  {Aspelmeyer},\ and\ \citenamefont {Painter}}]{chan2011}%
  \BibitemOpen
  \bibfield  {author} {\bibinfo {author} {\bibfnamefont {J.}~\bibnamefont
  {Chan}}, \bibinfo {author} {\bibfnamefont {T.~P.~M.}\ \bibnamefont {Alegre}},
  \bibinfo {author} {\bibfnamefont {A.~H.}\ \bibnamefont {{Safavi-Naeini}}},
  \bibinfo {author} {\bibfnamefont {J.~T.}\ \bibnamefont {Hill}}, \bibinfo
  {author} {\bibfnamefont {A.}~\bibnamefont {Krause}}, \bibinfo {author}
  {\bibfnamefont {S.}~\bibnamefont {Gr{\"o}blacher}}, \bibinfo {author}
  {\bibfnamefont {M.}~\bibnamefont {Aspelmeyer}},\ and\ \bibinfo {author}
  {\bibfnamefont {O.}~\bibnamefont {Painter}},\ }\bibfield  {title} {\bibinfo
  {title} {Laser cooling of a nanomechanical oscillator into its quantum ground
  state},\ }\href {https://doi.org/10.1038/nature10461} {\bibfield  {journal}
  {\bibinfo  {journal} {Nature}\ }\textbf {\bibinfo {volume} {478}},\ \bibinfo
  {pages} {89} (\bibinfo {year} {2011})}\BibitemShut {NoStop}%
\bibitem [{\citenamefont {Wollman}\ \emph {et~al.}(2015)\citenamefont
  {Wollman}, \citenamefont {Lei}, \citenamefont {Weinstein}, \citenamefont
  {Suh}, \citenamefont {Kronwald}, \citenamefont {Marquardt}, \citenamefont
  {Clerk},\ and\ \citenamefont {Schwab}}]{wollman2015}%
  \BibitemOpen
  \bibfield  {author} {\bibinfo {author} {\bibfnamefont {E.~E.}\ \bibnamefont
  {Wollman}}, \bibinfo {author} {\bibfnamefont {C.~U.}\ \bibnamefont {Lei}},
  \bibinfo {author} {\bibfnamefont {A.~J.}\ \bibnamefont {Weinstein}}, \bibinfo
  {author} {\bibfnamefont {J.}~\bibnamefont {Suh}}, \bibinfo {author}
  {\bibfnamefont {A.}~\bibnamefont {Kronwald}}, \bibinfo {author}
  {\bibfnamefont {F.}~\bibnamefont {Marquardt}}, \bibinfo {author}
  {\bibfnamefont {A.~A.}\ \bibnamefont {Clerk}},\ and\ \bibinfo {author}
  {\bibfnamefont {K.~C.}\ \bibnamefont {Schwab}},\ }\bibfield  {title}
  {\bibinfo {title} {Quantum squeezing of motion in a mechanical resonator},\
  }\href {https://doi.org/10.1126/science.aac5138} {\bibfield  {journal}
  {\bibinfo  {journal} {Science}\ }\textbf {\bibinfo {volume} {349}},\ \bibinfo
  {pages} {952} (\bibinfo {year} {2015})}\BibitemShut {NoStop}%
\bibitem [{\citenamefont {Pirkkalainen}\ \emph
  {et~al.}(2015{\natexlab{a}})\citenamefont {Pirkkalainen}, \citenamefont
  {Damsk{\"a}gg}, \citenamefont {Brandt}, \citenamefont {Massel},\ and\
  \citenamefont {Sillanp{\"a}{\"a}}}]{pirkkalainen2015}%
  \BibitemOpen
  \bibfield  {author} {\bibinfo {author} {\bibfnamefont {J.-M.}\ \bibnamefont
  {Pirkkalainen}}, \bibinfo {author} {\bibfnamefont {E.}~\bibnamefont
  {Damsk{\"a}gg}}, \bibinfo {author} {\bibfnamefont {M.}~\bibnamefont
  {Brandt}}, \bibinfo {author} {\bibfnamefont {F.}~\bibnamefont {Massel}},\
  and\ \bibinfo {author} {\bibfnamefont {M.~A.}\ \bibnamefont
  {Sillanp{\"a}{\"a}}},\ }\bibfield  {title} {\bibinfo {title} {Squeezing of
  {{Quantum Noise}} of {{Motion}} in a {{Micromechanical Resonator}}},\ }\href
  {https://doi.org/10.1103/PhysRevLett.115.243601} {\bibfield  {journal}
  {\bibinfo  {journal} {Physical Review Letters}\ }\textbf {\bibinfo {volume}
  {115}},\ \bibinfo {pages} {243601} (\bibinfo {year}
  {2015}{\natexlab{a}})}\BibitemShut {NoStop}%
\bibitem [{\citenamefont {Lecocq}\ \emph {et~al.}(2015)\citenamefont {Lecocq},
  \citenamefont {Clark}, \citenamefont {Simmonds}, \citenamefont {Aumentado},\
  and\ \citenamefont {Teufel}}]{lecocq2015}%
  \BibitemOpen
  \bibfield  {author} {\bibinfo {author} {\bibfnamefont {F.}~\bibnamefont
  {Lecocq}}, \bibinfo {author} {\bibfnamefont {J.~B.}\ \bibnamefont {Clark}},
  \bibinfo {author} {\bibfnamefont {R.~W.}\ \bibnamefont {Simmonds}}, \bibinfo
  {author} {\bibfnamefont {J.}~\bibnamefont {Aumentado}},\ and\ \bibinfo
  {author} {\bibfnamefont {J.~D.}\ \bibnamefont {Teufel}},\ }\bibfield  {title}
  {\bibinfo {title} {Quantum {{Nondemolition Measurement}} of a {{Nonclassical
  State}} of a {{Massive Object}}},\ }\href
  {https://doi.org/10.1103/PhysRevX.5.041037} {\bibfield  {journal} {\bibinfo
  {journal} {Physical Review X}\ }\textbf {\bibinfo {volume} {5}},\ \bibinfo
  {pages} {041037} (\bibinfo {year} {2015})}\BibitemShut {NoStop}%
\bibitem [{\citenamefont {Palomaki}\ \emph {et~al.}(2013)\citenamefont
  {Palomaki}, \citenamefont {Teufel}, \citenamefont {Simmonds},\ and\
  \citenamefont {Lehnert}}]{palomaki2013}%
  \BibitemOpen
  \bibfield  {author} {\bibinfo {author} {\bibfnamefont {T.~A.}\ \bibnamefont
  {Palomaki}}, \bibinfo {author} {\bibfnamefont {J.~D.}\ \bibnamefont
  {Teufel}}, \bibinfo {author} {\bibfnamefont {R.~W.}\ \bibnamefont
  {Simmonds}},\ and\ \bibinfo {author} {\bibfnamefont {K.~W.}\ \bibnamefont
  {Lehnert}},\ }\bibfield  {title} {\bibinfo {title} {Entangling {{Mechanical
  Motion}} with {{Microwave Fields}}},\ }\href
  {https://doi.org/10.1126/science.1244563} {\bibfield  {journal} {\bibinfo
  {journal} {Science}\ }\textbf {\bibinfo {volume} {342}},\ \bibinfo {pages}
  {710} (\bibinfo {year} {2013})}\BibitemShut {NoStop}%
\bibitem [{\citenamefont {{Ockeloen-Korppi}}\ \emph {et~al.}(2018)\citenamefont
  {{Ockeloen-Korppi}}, \citenamefont {Damsk{\"a}gg}, \citenamefont
  {Pirkkalainen}, \citenamefont {Asjad}, \citenamefont {Clerk}, \citenamefont
  {Massel}, \citenamefont {Woolley},\ and\ \citenamefont
  {Sillanp{\"a}{\"a}}}]{ockeloen-korppi2018}%
  \BibitemOpen
  \bibfield  {author} {\bibinfo {author} {\bibfnamefont {C.~F.}\ \bibnamefont
  {{Ockeloen-Korppi}}}, \bibinfo {author} {\bibfnamefont {E.}~\bibnamefont
  {Damsk{\"a}gg}}, \bibinfo {author} {\bibfnamefont {J.-M.}\ \bibnamefont
  {Pirkkalainen}}, \bibinfo {author} {\bibfnamefont {M.}~\bibnamefont {Asjad}},
  \bibinfo {author} {\bibfnamefont {A.~A.}\ \bibnamefont {Clerk}}, \bibinfo
  {author} {\bibfnamefont {F.}~\bibnamefont {Massel}}, \bibinfo {author}
  {\bibfnamefont {M.~J.}\ \bibnamefont {Woolley}},\ and\ \bibinfo {author}
  {\bibfnamefont {M.~A.}\ \bibnamefont {Sillanp{\"a}{\"a}}},\ }\bibfield
  {title} {\bibinfo {title} {Stabilized entanglement of massive mechanical
  oscillators},\ }\href {https://doi.org/10.1038/s41586-018-0038-x} {\bibfield
  {journal} {\bibinfo  {journal} {Nature}\ }\textbf {\bibinfo {volume} {556}},\
  \bibinfo {pages} {478} (\bibinfo {year} {2018})}\BibitemShut {NoStop}%
\bibitem [{\citenamefont {Riedinger}\ \emph {et~al.}(2018)\citenamefont
  {Riedinger}, \citenamefont {Wallucks}, \citenamefont {Marinkovi{\'c}},
  \citenamefont {L{\"o}schnauer}, \citenamefont {Aspelmeyer}, \citenamefont
  {Hong},\ and\ \citenamefont {Gr{\"o}blacher}}]{riedinger2018}%
  \BibitemOpen
  \bibfield  {author} {\bibinfo {author} {\bibfnamefont {R.}~\bibnamefont
  {Riedinger}}, \bibinfo {author} {\bibfnamefont {A.}~\bibnamefont {Wallucks}},
  \bibinfo {author} {\bibfnamefont {I.}~\bibnamefont {Marinkovi{\'c}}},
  \bibinfo {author} {\bibfnamefont {C.}~\bibnamefont {L{\"o}schnauer}},
  \bibinfo {author} {\bibfnamefont {M.}~\bibnamefont {Aspelmeyer}}, \bibinfo
  {author} {\bibfnamefont {S.}~\bibnamefont {Hong}},\ and\ \bibinfo {author}
  {\bibfnamefont {S.}~\bibnamefont {Gr{\"o}blacher}},\ }\bibfield  {title}
  {\bibinfo {title} {Remote quantum entanglement between two micromechanical
  oscillators},\ }\href {https://doi.org/10.1038/s41586-018-0036-z} {\bibfield
  {journal} {\bibinfo  {journal} {Nature}\ }\textbf {\bibinfo {volume} {556}},\
  \bibinfo {pages} {473} (\bibinfo {year} {2018})}\BibitemShut {NoStop}%
\bibitem [{\citenamefont {Kotler}\ \emph {et~al.}(2021)\citenamefont {Kotler},
  \citenamefont {Peterson}, \citenamefont {Shojaee}, \citenamefont {Lecocq},
  \citenamefont {Cicak}, \citenamefont {Kwiatkowski}, \citenamefont {Geller},
  \citenamefont {Glancy}, \citenamefont {Knill}, \citenamefont {Simmonds},
  \citenamefont {Aumentado},\ and\ \citenamefont {Teufel}}]{kotler2021}%
  \BibitemOpen
  \bibfield  {author} {\bibinfo {author} {\bibfnamefont {S.}~\bibnamefont
  {Kotler}}, \bibinfo {author} {\bibfnamefont {G.~A.}\ \bibnamefont
  {Peterson}}, \bibinfo {author} {\bibfnamefont {E.}~\bibnamefont {Shojaee}},
  \bibinfo {author} {\bibfnamefont {F.}~\bibnamefont {Lecocq}}, \bibinfo
  {author} {\bibfnamefont {K.}~\bibnamefont {Cicak}}, \bibinfo {author}
  {\bibfnamefont {A.}~\bibnamefont {Kwiatkowski}}, \bibinfo {author}
  {\bibfnamefont {S.}~\bibnamefont {Geller}}, \bibinfo {author} {\bibfnamefont
  {S.}~\bibnamefont {Glancy}}, \bibinfo {author} {\bibfnamefont
  {E.}~\bibnamefont {Knill}}, \bibinfo {author} {\bibfnamefont {R.~W.}\
  \bibnamefont {Simmonds}}, \bibinfo {author} {\bibfnamefont {J.}~\bibnamefont
  {Aumentado}},\ and\ \bibinfo {author} {\bibfnamefont {J.~D.}\ \bibnamefont
  {Teufel}},\ }\bibfield  {title} {\bibinfo {title} {Direct observation of
  deterministic macroscopic entanglement},\ }\href
  {https://doi.org/10.1126/science.abf2998} {\bibfield  {journal} {\bibinfo
  {journal} {Science}\ }\textbf {\bibinfo {volume} {372}},\ \bibinfo {pages}
  {622} (\bibinfo {year} {2021})}\BibitemShut {NoStop}%
\bibitem [{\citenamefont {Rabl}(2011)}]{rabl2011}%
  \BibitemOpen
  \bibfield  {author} {\bibinfo {author} {\bibfnamefont {P.}~\bibnamefont
  {Rabl}},\ }\bibfield  {title} {\bibinfo {title} {Photon {{Blockade Effect}}
  in {{Optomechanical Systems}}},\ }\href
  {https://doi.org/10.1103/PhysRevLett.107.063601} {\bibfield  {journal}
  {\bibinfo  {journal} {Physical Review Letters}\ }\textbf {\bibinfo {volume}
  {107}},\ \bibinfo {pages} {063601} (\bibinfo {year} {2011})}\BibitemShut
  {NoStop}%
\bibitem [{\citenamefont {Nunnenkamp}\ \emph {et~al.}(2011)\citenamefont
  {Nunnenkamp}, \citenamefont {B{\o}rkje},\ and\ \citenamefont
  {Girvin}}]{nunnenkamp2011}%
  \BibitemOpen
  \bibfield  {author} {\bibinfo {author} {\bibfnamefont {A.}~\bibnamefont
  {Nunnenkamp}}, \bibinfo {author} {\bibfnamefont {K.}~\bibnamefont
  {B{\o}rkje}},\ and\ \bibinfo {author} {\bibfnamefont {S.~M.}\ \bibnamefont
  {Girvin}},\ }\bibfield  {title} {\bibinfo {title} {Single-{{Photon
  Optomechanics}}},\ }\href {https://doi.org/10.1103/PhysRevLett.107.063602}
  {\bibfield  {journal} {\bibinfo  {journal} {Physical Review Letters}\
  }\textbf {\bibinfo {volume} {107}},\ \bibinfo {pages} {063602} (\bibinfo
  {year} {2011})}\BibitemShut {NoStop}%
\bibitem [{\citenamefont {Lemonde}\ \emph {et~al.}(2013)\citenamefont
  {Lemonde}, \citenamefont {Didier},\ and\ \citenamefont
  {Clerk}}]{lemonde2013}%
  \BibitemOpen
  \bibfield  {author} {\bibinfo {author} {\bibfnamefont {M.-A.}\ \bibnamefont
  {Lemonde}}, \bibinfo {author} {\bibfnamefont {N.}~\bibnamefont {Didier}},\
  and\ \bibinfo {author} {\bibfnamefont {A.~A.}\ \bibnamefont {Clerk}},\
  }\bibfield  {title} {\bibinfo {title} {Nonlinear {{Interaction Effects}} in a
  {{Strongly Driven Optomechanical Cavity}}},\ }\href
  {https://doi.org/10.1103/PhysRevLett.111.053602} {\bibfield  {journal}
  {\bibinfo  {journal} {Physical Review Letters}\ }\textbf {\bibinfo {volume}
  {111}},\ \bibinfo {pages} {053602} (\bibinfo {year} {2013})}\BibitemShut
  {NoStop}%
\bibitem [{\citenamefont {Qian}\ \emph {et~al.}(2012)\citenamefont {Qian},
  \citenamefont {Clerk}, \citenamefont {Hammerer},\ and\ \citenamefont
  {Marquardt}}]{qian2012}%
  \BibitemOpen
  \bibfield  {author} {\bibinfo {author} {\bibfnamefont {J.}~\bibnamefont
  {Qian}}, \bibinfo {author} {\bibfnamefont {A.~A.}\ \bibnamefont {Clerk}},
  \bibinfo {author} {\bibfnamefont {K.}~\bibnamefont {Hammerer}},\ and\
  \bibinfo {author} {\bibfnamefont {F.}~\bibnamefont {Marquardt}},\ }\bibfield
  {title} {\bibinfo {title} {Quantum {{Signatures}} of the {{Optomechanical
  Instability}}},\ }\href {https://doi.org/10.1103/PhysRevLett.109.253601}
  {\bibfield  {journal} {\bibinfo  {journal} {Physical Review Letters}\
  }\textbf {\bibinfo {volume} {109}},\ \bibinfo {pages} {253601} (\bibinfo
  {year} {2012})}\BibitemShut {NoStop}%
\bibitem [{\citenamefont {Nation}(2013)}]{nation2013}%
  \BibitemOpen
  \bibfield  {author} {\bibinfo {author} {\bibfnamefont {P.~D.}\ \bibnamefont
  {Nation}},\ }\bibfield  {title} {\bibinfo {title} {Nonclassical mechanical
  states in an optomechanical micromaser analog},\ }\href
  {https://doi.org/10.1103/PhysRevA.88.053828} {\bibfield  {journal} {\bibinfo
  {journal} {Physical Review A}\ }\textbf {\bibinfo {volume} {88}},\ \bibinfo
  {pages} {053828} (\bibinfo {year} {2013})}\BibitemShut {NoStop}%
\bibitem [{\citenamefont {L{\"o}rch}\ \emph {et~al.}(2014)\citenamefont
  {L{\"o}rch}, \citenamefont {Qian}, \citenamefont {Clerk}, \citenamefont
  {Marquardt},\ and\ \citenamefont {Hammerer}}]{lorch2014}%
  \BibitemOpen
  \bibfield  {author} {\bibinfo {author} {\bibfnamefont {N.}~\bibnamefont
  {L{\"o}rch}}, \bibinfo {author} {\bibfnamefont {J.}~\bibnamefont {Qian}},
  \bibinfo {author} {\bibfnamefont {A.}~\bibnamefont {Clerk}}, \bibinfo
  {author} {\bibfnamefont {F.}~\bibnamefont {Marquardt}},\ and\ \bibinfo
  {author} {\bibfnamefont {K.}~\bibnamefont {Hammerer}},\ }\bibfield  {title}
  {\bibinfo {title} {Laser {{Theory}} for {{Optomechanics}}: {{Limit Cycles}}
  in the {{Quantum Regime}}},\ }\href
  {https://doi.org/10.1103/PhysRevX.4.011015} {\bibfield  {journal} {\bibinfo
  {journal} {Physical Review X}\ }\textbf {\bibinfo {volume} {4}},\ \bibinfo
  {pages} {011015} (\bibinfo {year} {2014})}\BibitemShut {NoStop}%
\bibitem [{\citenamefont {Hauer}\ \emph {et~al.}(2023)\citenamefont {Hauer},
  \citenamefont {Combes},\ and\ \citenamefont {Teufel}}]{hauer2023}%
  \BibitemOpen
  \bibfield  {author} {\bibinfo {author} {\bibfnamefont {B.~D.}\ \bibnamefont
  {Hauer}}, \bibinfo {author} {\bibfnamefont {J.}~\bibnamefont {Combes}},\ and\
  \bibinfo {author} {\bibfnamefont {J.~D.}\ \bibnamefont {Teufel}},\ }\bibfield
   {title} {\bibinfo {title} {Nonlinear {{Sideband Cooling}} to a {{Cat State}}
  of {{Motion}}},\ }\href {https://doi.org/10.1103/PhysRevLett.130.213604}
  {\bibfield  {journal} {\bibinfo  {journal} {Physical Review Letters}\
  }\textbf {\bibinfo {volume} {130}},\ \bibinfo {pages} {213604} (\bibinfo
  {year} {2023})}\BibitemShut {NoStop}%
\bibitem [{\citenamefont {Mancini}\ \emph {et~al.}(1997)\citenamefont
  {Mancini}, \citenamefont {Man'ko},\ and\ \citenamefont
  {Tombesi}}]{mancini1997}%
  \BibitemOpen
  \bibfield  {author} {\bibinfo {author} {\bibfnamefont {S.}~\bibnamefont
  {Mancini}}, \bibinfo {author} {\bibfnamefont {V.~I.}\ \bibnamefont
  {Man'ko}},\ and\ \bibinfo {author} {\bibfnamefont {P.}~\bibnamefont
  {Tombesi}},\ }\bibfield  {title} {\bibinfo {title} {Ponderomotive control of
  quantum macroscopic coherence},\ }\href
  {https://doi.org/10.1103/PhysRevA.55.3042} {\bibfield  {journal} {\bibinfo
  {journal} {Physical Review A}\ }\textbf {\bibinfo {volume} {55}},\ \bibinfo
  {pages} {3042} (\bibinfo {year} {1997})}\BibitemShut {NoStop}%
\bibitem [{\citenamefont {Bose}\ \emph {et~al.}(1997)\citenamefont {Bose},
  \citenamefont {Jacobs},\ and\ \citenamefont {Knight}}]{bose1997}%
  \BibitemOpen
  \bibfield  {author} {\bibinfo {author} {\bibfnamefont {S.}~\bibnamefont
  {Bose}}, \bibinfo {author} {\bibfnamefont {K.}~\bibnamefont {Jacobs}},\ and\
  \bibinfo {author} {\bibfnamefont {P.~L.}\ \bibnamefont {Knight}},\ }\bibfield
   {title} {\bibinfo {title} {Preparation of nonclassical states in cavities
  with a moving mirror},\ }\href {https://doi.org/10.1103/PhysRevA.56.4175}
  {\bibfield  {journal} {\bibinfo  {journal} {Physical Review A}\ }\textbf
  {\bibinfo {volume} {56}},\ \bibinfo {pages} {4175} (\bibinfo {year}
  {1997})}\BibitemShut {NoStop}%
\bibitem [{\citenamefont {Yurke}\ and\ \citenamefont
  {Stoler}(1986)}]{yurke1986}%
  \BibitemOpen
  \bibfield  {author} {\bibinfo {author} {\bibfnamefont {B.}~\bibnamefont
  {Yurke}}\ and\ \bibinfo {author} {\bibfnamefont {D.}~\bibnamefont {Stoler}},\
  }\bibfield  {title} {\bibinfo {title} {Generating quantum mechanical
  superpositions of macroscopically distinguishable states via amplitude
  dispersion},\ }\href {https://doi.org/10.1103/PhysRevLett.57.13} {\bibfield
  {journal} {\bibinfo  {journal} {Physical Review Letters}\ }\textbf {\bibinfo
  {volume} {57}},\ \bibinfo {pages} {13} (\bibinfo {year} {1986})}\BibitemShut
  {NoStop}%
\bibitem [{\citenamefont {Bose}\ \emph {et~al.}(1999)\citenamefont {Bose},
  \citenamefont {Jacobs},\ and\ \citenamefont {Knight}}]{bose1999}%
  \BibitemOpen
  \bibfield  {author} {\bibinfo {author} {\bibfnamefont {S.}~\bibnamefont
  {Bose}}, \bibinfo {author} {\bibfnamefont {K.}~\bibnamefont {Jacobs}},\ and\
  \bibinfo {author} {\bibfnamefont {P.~L.}\ \bibnamefont {Knight}},\ }\bibfield
   {title} {\bibinfo {title} {Scheme to probe the decoherence of a macroscopic
  object},\ }\href {https://doi.org/10.1103/PhysRevA.59.3204} {\bibfield
  {journal} {\bibinfo  {journal} {Physical Review A}\ }\textbf {\bibinfo
  {volume} {59}},\ \bibinfo {pages} {3204} (\bibinfo {year}
  {1999})}\BibitemShut {NoStop}%
\bibitem [{\citenamefont {Marshall}\ \emph {et~al.}(2003)\citenamefont
  {Marshall}, \citenamefont {Simon}, \citenamefont {Penrose},\ and\
  \citenamefont {Bouwmeester}}]{marshall2003}%
  \BibitemOpen
  \bibfield  {author} {\bibinfo {author} {\bibfnamefont {W.}~\bibnamefont
  {Marshall}}, \bibinfo {author} {\bibfnamefont {C.}~\bibnamefont {Simon}},
  \bibinfo {author} {\bibfnamefont {R.}~\bibnamefont {Penrose}},\ and\ \bibinfo
  {author} {\bibfnamefont {D.}~\bibnamefont {Bouwmeester}},\ }\bibfield
  {title} {\bibinfo {title} {Towards {{Quantum Superpositions}} of a
  {{Mirror}}},\ }\href {https://doi.org/10.1103/PhysRevLett.91.130401}
  {\bibfield  {journal} {\bibinfo  {journal} {Physical Review Letters}\
  }\textbf {\bibinfo {volume} {91}},\ \bibinfo {pages} {130401} (\bibinfo
  {year} {2003})}\BibitemShut {NoStop}%
\bibitem [{\citenamefont {Pepper}\ \emph {et~al.}(2012)\citenamefont {Pepper},
  \citenamefont {Ghobadi}, \citenamefont {Jeffrey}, \citenamefont {Simon},\
  and\ \citenamefont {Bouwmeester}}]{pepper2012}%
  \BibitemOpen
  \bibfield  {author} {\bibinfo {author} {\bibfnamefont {B.}~\bibnamefont
  {Pepper}}, \bibinfo {author} {\bibfnamefont {R.}~\bibnamefont {Ghobadi}},
  \bibinfo {author} {\bibfnamefont {E.}~\bibnamefont {Jeffrey}}, \bibinfo
  {author} {\bibfnamefont {C.}~\bibnamefont {Simon}},\ and\ \bibinfo {author}
  {\bibfnamefont {D.}~\bibnamefont {Bouwmeester}},\ }\bibfield  {title}
  {\bibinfo {title} {Optomechanical {{Superpositions}} via {{Nested
  Interferometry}}},\ }\href {https://doi.org/10.1103/PhysRevLett.109.023601}
  {\bibfield  {journal} {\bibinfo  {journal} {Physical Review Letters}\
  }\textbf {\bibinfo {volume} {109}},\ \bibinfo {pages} {023601} (\bibinfo
  {year} {2012})}\BibitemShut {NoStop}%
\bibitem [{\citenamefont {Hong}\ \emph {et~al.}(2013)\citenamefont {Hong},
  \citenamefont {Yang}, \citenamefont {Miao},\ and\ \citenamefont
  {Chen}}]{hong2013}%
  \BibitemOpen
  \bibfield  {author} {\bibinfo {author} {\bibfnamefont {T.}~\bibnamefont
  {Hong}}, \bibinfo {author} {\bibfnamefont {H.}~\bibnamefont {Yang}}, \bibinfo
  {author} {\bibfnamefont {H.}~\bibnamefont {Miao}},\ and\ \bibinfo {author}
  {\bibfnamefont {Y.}~\bibnamefont {Chen}},\ }\bibfield  {title} {\bibinfo
  {title} {Open quantum dynamics of single-photon optomechanical devices},\
  }\href {https://doi.org/10.1103/PhysRevA.88.023812} {\bibfield  {journal}
  {\bibinfo  {journal} {Physical Review A}\ }\textbf {\bibinfo {volume} {88}},\
  \bibinfo {pages} {023812} (\bibinfo {year} {2013})}\BibitemShut {NoStop}%
\bibitem [{\citenamefont {Akram}\ \emph {et~al.}(2013)\citenamefont {Akram},
  \citenamefont {Bowen},\ and\ \citenamefont {Milburn}}]{akram2013}%
  \BibitemOpen
  \bibfield  {author} {\bibinfo {author} {\bibfnamefont {U.}~\bibnamefont
  {Akram}}, \bibinfo {author} {\bibfnamefont {W.~P.}\ \bibnamefont {Bowen}},\
  and\ \bibinfo {author} {\bibfnamefont {G.~J.}\ \bibnamefont {Milburn}},\
  }\bibfield  {title} {\bibinfo {title} {Entangled mechanical cat states via
  conditional single photon optomechanics},\ }\href
  {https://doi.org/10.1088/1367-2630/15/9/093007} {\bibfield  {journal}
  {\bibinfo  {journal} {New Journal of Physics}\ }\textbf {\bibinfo {volume}
  {15}},\ \bibinfo {pages} {093007} (\bibinfo {year} {2013})}\BibitemShut
  {NoStop}%
\bibitem [{\citenamefont {Mancini}\ and\ \citenamefont
  {Tombesi}(1994)}]{mancini1994}%
  \BibitemOpen
  \bibfield  {author} {\bibinfo {author} {\bibfnamefont {S.}~\bibnamefont
  {Mancini}}\ and\ \bibinfo {author} {\bibfnamefont {P.}~\bibnamefont
  {Tombesi}},\ }\bibfield  {title} {\bibinfo {title} {Quantum noise reduction
  by radiation pressure},\ }\href {https://doi.org/10.1103/PhysRevA.49.4055}
  {\bibfield  {journal} {\bibinfo  {journal} {Physical Review A}\ }\textbf
  {\bibinfo {volume} {49}},\ \bibinfo {pages} {4055} (\bibinfo {year}
  {1994})}\BibitemShut {NoStop}%
\bibitem [{\citenamefont {Law}(1995)}]{law1995}%
  \BibitemOpen
  \bibfield  {author} {\bibinfo {author} {\bibfnamefont {C.~K.}\ \bibnamefont
  {Law}},\ }\bibfield  {title} {\bibinfo {title} {Interaction between a moving
  mirror and radiation pressure: {{A Hamiltonian}} formulation},\ }\href
  {https://doi.org/10.1103/PhysRevA.51.2537} {\bibfield  {journal} {\bibinfo
  {journal} {Physical Review A}\ }\textbf {\bibinfo {volume} {51}},\ \bibinfo
  {pages} {2537} (\bibinfo {year} {1995})}\BibitemShut {NoStop}%
\bibitem [{\citenamefont {Lang}\ and\ \citenamefont {Firsov}(1963)}]{lang1963}%
  \BibitemOpen
  \bibfield  {author} {\bibinfo {author} {\bibfnamefont {{\relax
  IG}.}~\bibnamefont {Lang}}\ and\ \bibinfo {author} {\bibfnamefont {Y.~A.}\
  \bibnamefont {Firsov}},\ }\bibfield  {title} {\bibinfo {title} {Kinetic
  theory of semiconductors with low mobility},\ }\href@noop {} {\bibfield
  {journal} {\bibinfo  {journal} {Soviet Physics--JETP [translation of Zhurnal
  Eksperimentalnoi i Teoreticheskoi Fiziki]}\ }\textbf {\bibinfo {volume}
  {16}},\ \bibinfo {pages} {1301} (\bibinfo {year} {1963})}\BibitemShut
  {NoStop}%
\bibitem [{\citenamefont {Mahan}(2000)}]{mahan2000}%
  \BibitemOpen
  \bibfield  {author} {\bibinfo {author} {\bibfnamefont {G.~D.}\ \bibnamefont
  {Mahan}},\ }\href@noop {} {\emph {\bibinfo {title} {Many-Particle Physics}}}\
  (\bibinfo  {publisher} {Springer Science \& Business Media},\ \bibinfo {year}
  {2000})\BibitemShut {NoStop}%
\bibitem [{\citenamefont {Marquardt}\ \emph {et~al.}(2007)\citenamefont
  {Marquardt}, \citenamefont {Chen}, \citenamefont {Clerk},\ and\ \citenamefont
  {Girvin}}]{marquardt2007}%
  \BibitemOpen
  \bibfield  {author} {\bibinfo {author} {\bibfnamefont {F.}~\bibnamefont
  {Marquardt}}, \bibinfo {author} {\bibfnamefont {J.~P.}\ \bibnamefont {Chen}},
  \bibinfo {author} {\bibfnamefont {A.~A.}\ \bibnamefont {Clerk}},\ and\
  \bibinfo {author} {\bibfnamefont {S.~M.}\ \bibnamefont {Girvin}},\ }\bibfield
   {title} {\bibinfo {title} {Quantum {{Theory}} of {{Cavity-Assisted Sideband
  Cooling}} of {{Mechanical Motion}}},\ }\href
  {https://doi.org/10.1103/PhysRevLett.99.093902} {\bibfield  {journal}
  {\bibinfo  {journal} {Physical Review Letters}\ }\textbf {\bibinfo {volume}
  {99}},\ \bibinfo {pages} {093902} (\bibinfo {year} {2007})}\BibitemShut
  {NoStop}%
\bibitem [{\citenamefont {Kirchmair}\ \emph {et~al.}(2013)\citenamefont
  {Kirchmair}, \citenamefont {Vlastakis}, \citenamefont {Leghtas},
  \citenamefont {Nigg}, \citenamefont {Paik}, \citenamefont {Ginossar},
  \citenamefont {Mirrahimi}, \citenamefont {Frunzio}, \citenamefont {Girvin},\
  and\ \citenamefont {Schoelkopf}}]{kirchmair2013}%
  \BibitemOpen
  \bibfield  {author} {\bibinfo {author} {\bibfnamefont {G.}~\bibnamefont
  {Kirchmair}}, \bibinfo {author} {\bibfnamefont {B.}~\bibnamefont
  {Vlastakis}}, \bibinfo {author} {\bibfnamefont {Z.}~\bibnamefont {Leghtas}},
  \bibinfo {author} {\bibfnamefont {S.~E.}\ \bibnamefont {Nigg}}, \bibinfo
  {author} {\bibfnamefont {H.}~\bibnamefont {Paik}}, \bibinfo {author}
  {\bibfnamefont {E.}~\bibnamefont {Ginossar}}, \bibinfo {author}
  {\bibfnamefont {M.}~\bibnamefont {Mirrahimi}}, \bibinfo {author}
  {\bibfnamefont {L.}~\bibnamefont {Frunzio}}, \bibinfo {author} {\bibfnamefont
  {S.~M.}\ \bibnamefont {Girvin}},\ and\ \bibinfo {author} {\bibfnamefont
  {R.~J.}\ \bibnamefont {Schoelkopf}},\ }\bibfield  {title} {\bibinfo {title}
  {Observation of quantum state collapse and revival due to the single-photon
  {{Kerr}} effect},\ }\href {https://doi.org/10.1038/nature11902} {\bibfield
  {journal} {\bibinfo  {journal} {Nature}\ }\textbf {\bibinfo {volume} {495}},\
  \bibinfo {pages} {205} (\bibinfo {year} {2013})}\BibitemShut {NoStop}%
\bibitem [{\citenamefont {{Moya-Cessa}}\ and\ \citenamefont
  {Knight}(1993)}]{moya-cessa1993}%
  \BibitemOpen
  \bibfield  {author} {\bibinfo {author} {\bibfnamefont {H.}~\bibnamefont
  {{Moya-Cessa}}}\ and\ \bibinfo {author} {\bibfnamefont {P.~L.}\ \bibnamefont
  {Knight}},\ }\bibfield  {title} {\bibinfo {title} {Series representation of
  quantum-field quasiprobabilities},\ }\href
  {https://doi.org/10.1103/PhysRevA.48.2479} {\bibfield  {journal} {\bibinfo
  {journal} {Physical Review A}\ }\textbf {\bibinfo {volume} {48}},\ \bibinfo
  {pages} {2479} (\bibinfo {year} {1993})}\BibitemShut {NoStop}%
\bibitem [{\citenamefont {{de Oliveira}}\ \emph {et~al.}(1990)\citenamefont
  {{de Oliveira}}, \citenamefont {Kim}, \citenamefont {Knight},\ and\
  \citenamefont {Buek}}]{deoliveira1990}%
  \BibitemOpen
  \bibfield  {author} {\bibinfo {author} {\bibfnamefont {F.~A.~M.}\
  \bibnamefont {{de Oliveira}}}, \bibinfo {author} {\bibfnamefont {M.~S.}\
  \bibnamefont {Kim}}, \bibinfo {author} {\bibfnamefont {P.~L.}\ \bibnamefont
  {Knight}},\ and\ \bibinfo {author} {\bibfnamefont {V.}~\bibnamefont {Buek}},\
  }\bibfield  {title} {\bibinfo {title} {Properties of displaced number
  states},\ }\href {https://doi.org/10.1103/PhysRevA.41.2645} {\bibfield
  {journal} {\bibinfo  {journal} {Physical Review A}\ }\textbf {\bibinfo
  {volume} {41}},\ \bibinfo {pages} {2645} (\bibinfo {year}
  {1990})}\BibitemShut {NoStop}%
\bibitem [{\citenamefont {Wunsche}(1991)}]{wunsche1991}%
  \BibitemOpen
  \bibfield  {author} {\bibinfo {author} {\bibfnamefont {A.}~\bibnamefont
  {Wunsche}},\ }\bibfield  {title} {\bibinfo {title} {Displaced {{Fock}} states
  and their connection to quasiprobabilities},\ }\href
  {https://doi.org/10.1088/0954-8998/3/6/005} {\bibfield  {journal} {\bibinfo
  {journal} {Quantum Optics: Journal of the European Optical Society Part B}\
  }\textbf {\bibinfo {volume} {3}},\ \bibinfo {pages} {359} (\bibinfo {year}
  {1991})}\BibitemShut {NoStop}%
\bibitem [{\citenamefont {Ludwig}\ \emph {et~al.}(2008)\citenamefont {Ludwig},
  \citenamefont {Kubala},\ and\ \citenamefont {Marquardt}}]{ludwig2008}%
  \BibitemOpen
  \bibfield  {author} {\bibinfo {author} {\bibfnamefont {M.}~\bibnamefont
  {Ludwig}}, \bibinfo {author} {\bibfnamefont {B.}~\bibnamefont {Kubala}},\
  and\ \bibinfo {author} {\bibfnamefont {F.}~\bibnamefont {Marquardt}},\
  }\bibfield  {title} {\bibinfo {title} {The optomechanical instability in the
  quantum regime},\ }\href {https://doi.org/10.1088/1367-2630/10/9/095013}
  {\bibfield  {journal} {\bibinfo  {journal} {New Journal of Physics}\ }\textbf
  {\bibinfo {volume} {10}},\ \bibinfo {pages} {095013} (\bibinfo {year}
  {2008})}\BibitemShut {NoStop}%
\bibitem [{\citenamefont {Gardiner}\ and\ \citenamefont
  {Zoller}(2004)}]{gardiner2004}%
  \BibitemOpen
  \bibfield  {author} {\bibinfo {author} {\bibfnamefont {C.}~\bibnamefont
  {Gardiner}}\ and\ \bibinfo {author} {\bibfnamefont {P.}~\bibnamefont
  {Zoller}},\ }\href@noop {} {\emph {\bibinfo {title} {Quantum Noise: {{A}}
  Handbook of Markovian and Non-Markovian Quantum Stochastic Methods with
  Applications to Quantum Optics}}},\ Springer Series in Synergetics\ (\bibinfo
   {publisher} {Springer},\ \bibinfo {year} {2004})\BibitemShut {NoStop}%
\bibitem [{\citenamefont {Johansson}\ \emph {et~al.}(2012)\citenamefont
  {Johansson}, \citenamefont {Nation},\ and\ \citenamefont
  {Nori}}]{johansson2012}%
  \BibitemOpen
  \bibfield  {author} {\bibinfo {author} {\bibfnamefont {J.~R.}\ \bibnamefont
  {Johansson}}, \bibinfo {author} {\bibfnamefont {P.~D.}\ \bibnamefont
  {Nation}},\ and\ \bibinfo {author} {\bibfnamefont {F.}~\bibnamefont {Nori}},\
  }\bibfield  {title} {\bibinfo {title} {{{QuTiP}}: {{An}} open-source
  {{Python}} framework for the dynamics of open quantum systems},\ }\href
  {https://doi.org/10.1016/j.cpc.2012.02.021} {\bibfield  {journal} {\bibinfo
  {journal} {Computer Physics Communications}\ }\textbf {\bibinfo {volume}
  {183}},\ \bibinfo {pages} {1760} (\bibinfo {year} {2012})}\BibitemShut
  {NoStop}%
\bibitem [{\citenamefont {Johansson}\ \emph {et~al.}(2013)\citenamefont
  {Johansson}, \citenamefont {Nation},\ and\ \citenamefont
  {Nori}}]{johansson2013}%
  \BibitemOpen
  \bibfield  {author} {\bibinfo {author} {\bibfnamefont {J.~R.}\ \bibnamefont
  {Johansson}}, \bibinfo {author} {\bibfnamefont {P.~D.}\ \bibnamefont
  {Nation}},\ and\ \bibinfo {author} {\bibfnamefont {F.}~\bibnamefont {Nori}},\
  }\bibfield  {title} {\bibinfo {title} {{{QuTiP}} 2: {{A Python}} framework
  for the dynamics of open quantum systems},\ }\href
  {https://doi.org/10.1016/j.cpc.2012.11.019} {\bibfield  {journal} {\bibinfo
  {journal} {Computer Physics Communications}\ }\textbf {\bibinfo {volume}
  {184}},\ \bibinfo {pages} {1234} (\bibinfo {year} {2013})}\BibitemShut
  {NoStop}%
\bibitem [{\citenamefont {Fogliano}\ \emph {et~al.}(2021)\citenamefont
  {Fogliano}, \citenamefont {Besga}, \citenamefont {Reigue}, \citenamefont
  {Heringlake}, \citenamefont {{Mercier de L{\'e}pinay}}, \citenamefont
  {Vaneph}, \citenamefont {Reichel}, \citenamefont {Pigeau},\ and\
  \citenamefont {Arcizet}}]{fogliano2021}%
  \BibitemOpen
  \bibfield  {author} {\bibinfo {author} {\bibfnamefont {F.}~\bibnamefont
  {Fogliano}}, \bibinfo {author} {\bibfnamefont {B.}~\bibnamefont {Besga}},
  \bibinfo {author} {\bibfnamefont {A.}~\bibnamefont {Reigue}}, \bibinfo
  {author} {\bibfnamefont {P.}~\bibnamefont {Heringlake}}, \bibinfo {author}
  {\bibfnamefont {L.}~\bibnamefont {{Mercier de L{\'e}pinay}}}, \bibinfo
  {author} {\bibfnamefont {C.}~\bibnamefont {Vaneph}}, \bibinfo {author}
  {\bibfnamefont {J.}~\bibnamefont {Reichel}}, \bibinfo {author} {\bibfnamefont
  {B.}~\bibnamefont {Pigeau}},\ and\ \bibinfo {author} {\bibfnamefont
  {O.}~\bibnamefont {Arcizet}},\ }\bibfield  {title} {\bibinfo {title} {Mapping
  the {{Cavity Optomechanical Interaction}} with {{Subwavelength-Sized
  Ultrasensitive Nanomechanical Force Sensors}}},\ }\href
  {https://doi.org/10.1103/PhysRevX.11.021009} {\bibfield  {journal} {\bibinfo
  {journal} {Physical Review X}\ }\textbf {\bibinfo {volume} {11}},\ \bibinfo
  {pages} {021009} (\bibinfo {year} {2021})}\BibitemShut {NoStop}%
\bibitem [{\citenamefont {Chan}\ \emph {et~al.}(2012)\citenamefont {Chan},
  \citenamefont {{Safavi-Naeini}}, \citenamefont {Hill}, \citenamefont
  {Meenehan},\ and\ \citenamefont {Painter}}]{chan2012}%
  \BibitemOpen
  \bibfield  {author} {\bibinfo {author} {\bibfnamefont {J.}~\bibnamefont
  {Chan}}, \bibinfo {author} {\bibfnamefont {A.~H.}\ \bibnamefont
  {{Safavi-Naeini}}}, \bibinfo {author} {\bibfnamefont {J.~T.}\ \bibnamefont
  {Hill}}, \bibinfo {author} {\bibfnamefont {S.}~\bibnamefont {Meenehan}},\
  and\ \bibinfo {author} {\bibfnamefont {O.}~\bibnamefont {Painter}},\
  }\bibfield  {title} {\bibinfo {title} {Optimized optomechanical crystal
  cavity with acoustic radiation shield},\ }\href
  {https://doi.org/10.1063/1.4747726} {\bibfield  {journal} {\bibinfo
  {journal} {Applied Physics Letters}\ }\textbf {\bibinfo {volume} {101}},\
  \bibinfo {pages} {081115} (\bibinfo {year} {2012})}\BibitemShut {NoStop}%
\bibitem [{\citenamefont {Leijssen}\ and\ \citenamefont
  {Verhagen}(2015)}]{leijssen2015}%
  \BibitemOpen
  \bibfield  {author} {\bibinfo {author} {\bibfnamefont {R.}~\bibnamefont
  {Leijssen}}\ and\ \bibinfo {author} {\bibfnamefont {E.}~\bibnamefont
  {Verhagen}},\ }\bibfield  {title} {\bibinfo {title} {Strong optomechanical
  interactions in a sliced photonic crystal nanobeam},\ }\href
  {https://doi.org/10.1038/srep15974} {\bibfield  {journal} {\bibinfo
  {journal} {Scientific Reports}\ }\textbf {\bibinfo {volume} {5}},\ \bibinfo
  {pages} {15974} (\bibinfo {year} {2015})}\BibitemShut {NoStop}%
\bibitem [{\citenamefont {Bozkurt}\ \emph {et~al.}(2022)\citenamefont
  {Bozkurt}, \citenamefont {Joshi},\ and\ \citenamefont
  {Mirhosseini}}]{bozkurt2022}%
  \BibitemOpen
  \bibfield  {author} {\bibinfo {author} {\bibfnamefont {A.}~\bibnamefont
  {Bozkurt}}, \bibinfo {author} {\bibfnamefont {C.}~\bibnamefont {Joshi}},\
  and\ \bibinfo {author} {\bibfnamefont {M.}~\bibnamefont {Mirhosseini}},\
  }\bibfield  {title} {\bibinfo {title} {Deep sub-wavelength localization of
  light and sound in dielectric resonators},\ }\href
  {https://doi.org/10.1364/OE.455248} {\bibfield  {journal} {\bibinfo
  {journal} {Optics Express}\ }\textbf {\bibinfo {volume} {30}},\ \bibinfo
  {pages} {12378} (\bibinfo {year} {2022})}\BibitemShut {NoStop}%
\bibitem [{\citenamefont {Reinhardt}\ \emph {et~al.}(2016)\citenamefont
  {Reinhardt}, \citenamefont {M{\"u}ller}, \citenamefont {Bourassa},\ and\
  \citenamefont {Sankey}}]{reinhardt2016}%
  \BibitemOpen
  \bibfield  {author} {\bibinfo {author} {\bibfnamefont {C.}~\bibnamefont
  {Reinhardt}}, \bibinfo {author} {\bibfnamefont {T.}~\bibnamefont
  {M{\"u}ller}}, \bibinfo {author} {\bibfnamefont {A.}~\bibnamefont
  {Bourassa}},\ and\ \bibinfo {author} {\bibfnamefont {J.~C.}\ \bibnamefont
  {Sankey}},\ }\bibfield  {title} {\bibinfo {title} {Ultralow-{{Noise SiN
  Trampoline Resonators}} for {{Sensing}} and {{Optomechanics}}},\ }\href
  {https://doi.org/10.1103/PhysRevX.6.021001} {\bibfield  {journal} {\bibinfo
  {journal} {Physical Review X}\ }\textbf {\bibinfo {volume} {6}},\ \bibinfo
  {pages} {021001} (\bibinfo {year} {2016})}\BibitemShut {NoStop}%
\bibitem [{\citenamefont {Teufel}\ \emph
  {et~al.}(2011{\natexlab{b}})\citenamefont {Teufel}, \citenamefont {Li},
  \citenamefont {Allman}, \citenamefont {Cicak}, \citenamefont {Sirois},
  \citenamefont {Whittaker},\ and\ \citenamefont {Simmonds}}]{teufel2011b}%
  \BibitemOpen
  \bibfield  {author} {\bibinfo {author} {\bibfnamefont {J.~D.}\ \bibnamefont
  {Teufel}}, \bibinfo {author} {\bibfnamefont {D.}~\bibnamefont {Li}}, \bibinfo
  {author} {\bibfnamefont {M.~S.}\ \bibnamefont {Allman}}, \bibinfo {author}
  {\bibfnamefont {K.}~\bibnamefont {Cicak}}, \bibinfo {author} {\bibfnamefont
  {A.~J.}\ \bibnamefont {Sirois}}, \bibinfo {author} {\bibfnamefont {J.~D.}\
  \bibnamefont {Whittaker}},\ and\ \bibinfo {author} {\bibfnamefont {R.~W.}\
  \bibnamefont {Simmonds}},\ }\bibfield  {title} {\bibinfo {title} {Circuit
  cavity electromechanics in the strong-coupling regime},\ }\href
  {https://doi.org/10.1038/nature09898} {\bibfield  {journal} {\bibinfo
  {journal} {Nature}\ }\textbf {\bibinfo {volume} {471}},\ \bibinfo {pages}
  {204} (\bibinfo {year} {2011}{\natexlab{b}})}\BibitemShut {NoStop}%
\bibitem [{\citenamefont {Samanta}\ \emph {et~al.}(2023)\citenamefont
  {Samanta}, \citenamefont {De~Bonis}, \citenamefont {M{\o}ller}, \citenamefont
  {{Tormo-Queralt}}, \citenamefont {Yang}, \citenamefont {Urgell},
  \citenamefont {Stamenic}, \citenamefont {Thibeault}, \citenamefont {Jin},
  \citenamefont {Czaplewski}, \citenamefont {Pistolesi},\ and\ \citenamefont
  {Bachtold}}]{samanta2023}%
  \BibitemOpen
  \bibfield  {author} {\bibinfo {author} {\bibfnamefont {C.}~\bibnamefont
  {Samanta}}, \bibinfo {author} {\bibfnamefont {S.~L.}\ \bibnamefont
  {De~Bonis}}, \bibinfo {author} {\bibfnamefont {C.~B.}\ \bibnamefont
  {M{\o}ller}}, \bibinfo {author} {\bibfnamefont {R.}~\bibnamefont
  {{Tormo-Queralt}}}, \bibinfo {author} {\bibfnamefont {W.}~\bibnamefont
  {Yang}}, \bibinfo {author} {\bibfnamefont {C.}~\bibnamefont {Urgell}},
  \bibinfo {author} {\bibfnamefont {B.}~\bibnamefont {Stamenic}}, \bibinfo
  {author} {\bibfnamefont {B.}~\bibnamefont {Thibeault}}, \bibinfo {author}
  {\bibfnamefont {Y.}~\bibnamefont {Jin}}, \bibinfo {author} {\bibfnamefont
  {D.~A.}\ \bibnamefont {Czaplewski}}, \bibinfo {author} {\bibfnamefont
  {F.}~\bibnamefont {Pistolesi}},\ and\ \bibinfo {author} {\bibfnamefont
  {A.}~\bibnamefont {Bachtold}},\ }\bibfield  {title} {\bibinfo {title}
  {Nonlinear nanomechanical resonators approaching the quantum ground state},\
  }\href {https://doi.org/10.1038/s41567-023-02065-9} {\bibfield  {journal}
  {\bibinfo  {journal} {Nature Physics}\ }\textbf {\bibinfo {volume} {19}},\
  \bibinfo {pages} {1340} (\bibinfo {year} {2023})}\BibitemShut {NoStop}%
\bibitem [{\citenamefont {Heikkil{\"a}}\ \emph {et~al.}(2014)\citenamefont
  {Heikkil{\"a}}, \citenamefont {Massel}, \citenamefont {Tuorila},
  \citenamefont {Khan},\ and\ \citenamefont
  {Sillanp{\"a}{\"a}}}]{heikkila2014}%
  \BibitemOpen
  \bibfield  {author} {\bibinfo {author} {\bibfnamefont {T.~T.}\ \bibnamefont
  {Heikkil{\"a}}}, \bibinfo {author} {\bibfnamefont {F.}~\bibnamefont
  {Massel}}, \bibinfo {author} {\bibfnamefont {J.}~\bibnamefont {Tuorila}},
  \bibinfo {author} {\bibfnamefont {R.}~\bibnamefont {Khan}},\ and\ \bibinfo
  {author} {\bibfnamefont {M.~A.}\ \bibnamefont {Sillanp{\"a}{\"a}}},\
  }\bibfield  {title} {\bibinfo {title} {Enhancing {{Optomechanical Coupling}}
  via the {{Josephson Effect}}},\ }\href
  {https://doi.org/10.1103/PhysRevLett.112.203603} {\bibfield  {journal}
  {\bibinfo  {journal} {Physical Review Letters}\ }\textbf {\bibinfo {volume}
  {112}},\ \bibinfo {pages} {203603} (\bibinfo {year} {2014})}\BibitemShut
  {NoStop}%
\bibitem [{\citenamefont {Pirkkalainen}\ \emph
  {et~al.}(2015{\natexlab{b}})\citenamefont {Pirkkalainen}, \citenamefont
  {Cho}, \citenamefont {Massel}, \citenamefont {Tuorila}, \citenamefont
  {Heikkil{\"a}}, \citenamefont {Hakonen},\ and\ \citenamefont
  {Sillanp{\"a}{\"a}}}]{pirkkalainen2015a}%
  \BibitemOpen
  \bibfield  {author} {\bibinfo {author} {\bibfnamefont {J.-M.}\ \bibnamefont
  {Pirkkalainen}}, \bibinfo {author} {\bibfnamefont {S.~U.}\ \bibnamefont
  {Cho}}, \bibinfo {author} {\bibfnamefont {F.}~\bibnamefont {Massel}},
  \bibinfo {author} {\bibfnamefont {J.}~\bibnamefont {Tuorila}}, \bibinfo
  {author} {\bibfnamefont {T.~T.}\ \bibnamefont {Heikkil{\"a}}}, \bibinfo
  {author} {\bibfnamefont {P.~J.}\ \bibnamefont {Hakonen}},\ and\ \bibinfo
  {author} {\bibfnamefont {M.~A.}\ \bibnamefont {Sillanp{\"a}{\"a}}},\
  }\bibfield  {title} {\bibinfo {title} {Cavity optomechanics mediated by a
  quantum two-level system},\ }\href {https://doi.org/10.1038/ncomms7981}
  {\bibfield  {journal} {\bibinfo  {journal} {Nature Communications}\ }\textbf
  {\bibinfo {volume} {6}},\ \bibinfo {pages} {6981} (\bibinfo {year}
  {2015}{\natexlab{b}})}\BibitemShut {NoStop}%
\bibitem [{\citenamefont {Manninen}\ \emph {et~al.}(2022)\citenamefont
  {Manninen}, \citenamefont {Haque}, \citenamefont {Vitali},\ and\
  \citenamefont {Hakonen}}]{manninen2022}%
  \BibitemOpen
  \bibfield  {author} {\bibinfo {author} {\bibfnamefont {J.}~\bibnamefont
  {Manninen}}, \bibinfo {author} {\bibfnamefont {M.~T.}\ \bibnamefont {Haque}},
  \bibinfo {author} {\bibfnamefont {D.}~\bibnamefont {Vitali}},\ and\ \bibinfo
  {author} {\bibfnamefont {P.}~\bibnamefont {Hakonen}},\ }\bibfield  {title}
  {\bibinfo {title} {Enhancement of the optomechanical coupling and {{Kerr}}
  nonlinearity using the {{Josephson}} capacitance of a {{Cooper-pair}} box},\
  }\href {https://doi.org/10.1103/PhysRevB.105.144508} {\bibfield  {journal}
  {\bibinfo  {journal} {Physical Review B}\ }\textbf {\bibinfo {volume}
  {105}},\ \bibinfo {pages} {144508} (\bibinfo {year} {2022})}\BibitemShut
  {NoStop}%
\bibitem [{\citenamefont {Vanner}\ \emph {et~al.}(2015)\citenamefont {Vanner},
  \citenamefont {Pikovski},\ and\ \citenamefont {Kim}}]{vanner2015}%
  \BibitemOpen
  \bibfield  {author} {\bibinfo {author} {\bibfnamefont {M.~R.}\ \bibnamefont
  {Vanner}}, \bibinfo {author} {\bibfnamefont {I.}~\bibnamefont {Pikovski}},\
  and\ \bibinfo {author} {\bibfnamefont {M.~S.}\ \bibnamefont {Kim}},\
  }\bibfield  {title} {\bibinfo {title} {Towards optomechanical quantum state
  reconstruction of mechanical motion},\ }\href
  {https://doi.org/10.1002/andp.201400124} {\bibfield  {journal} {\bibinfo
  {journal} {Annalen der Physik}\ }\textbf {\bibinfo {volume} {527}},\ \bibinfo
  {pages} {15} (\bibinfo {year} {2015})}\BibitemShut {NoStop}%
\bibitem [{\citenamefont {Suh}\ \emph {et~al.}(2014)\citenamefont {Suh},
  \citenamefont {Weinstein}, \citenamefont {Lei}, \citenamefont {Wollman},
  \citenamefont {Steinke}, \citenamefont {Meystre}, \citenamefont {Clerk},\
  and\ \citenamefont {Schwab}}]{suh2014}%
  \BibitemOpen
  \bibfield  {author} {\bibinfo {author} {\bibfnamefont {J.}~\bibnamefont
  {Suh}}, \bibinfo {author} {\bibfnamefont {A.~J.}\ \bibnamefont {Weinstein}},
  \bibinfo {author} {\bibfnamefont {C.~U.}\ \bibnamefont {Lei}}, \bibinfo
  {author} {\bibfnamefont {E.~E.}\ \bibnamefont {Wollman}}, \bibinfo {author}
  {\bibfnamefont {S.~K.}\ \bibnamefont {Steinke}}, \bibinfo {author}
  {\bibfnamefont {P.}~\bibnamefont {Meystre}}, \bibinfo {author} {\bibfnamefont
  {A.~A.}\ \bibnamefont {Clerk}},\ and\ \bibinfo {author} {\bibfnamefont
  {K.~C.}\ \bibnamefont {Schwab}},\ }\bibfield  {title} {\bibinfo {title}
  {Mechanically detecting and avoiding the quantum fluctuations of a microwave
  field},\ }\href {https://doi.org/10.1126/science.1253258} {\bibfield
  {journal} {\bibinfo  {journal} {Science}\ }\textbf {\bibinfo {volume}
  {344}},\ \bibinfo {pages} {1262} (\bibinfo {year} {2014})}\BibitemShut
  {NoStop}%
\bibitem [{\citenamefont {Delaney}\ \emph {et~al.}(2019)\citenamefont
  {Delaney}, \citenamefont {Reed}, \citenamefont {Andrews},\ and\ \citenamefont
  {Lehnert}}]{delaney2019}%
  \BibitemOpen
  \bibfield  {author} {\bibinfo {author} {\bibfnamefont {R.~D.}\ \bibnamefont
  {Delaney}}, \bibinfo {author} {\bibfnamefont {A.~P.}\ \bibnamefont {Reed}},
  \bibinfo {author} {\bibfnamefont {R.~W.}\ \bibnamefont {Andrews}},\ and\
  \bibinfo {author} {\bibfnamefont {K.~W.}\ \bibnamefont {Lehnert}},\
  }\bibfield  {title} {\bibinfo {title} {Measurement of {{Motion}} beyond the
  {{Quantum Limit}} by {{Transient Amplification}}},\ }\href
  {https://doi.org/10.1103/PhysRevLett.123.183603} {\bibfield  {journal}
  {\bibinfo  {journal} {Physical Review Letters}\ }\textbf {\bibinfo {volume}
  {123}},\ \bibinfo {pages} {183603} (\bibinfo {year} {2019})}\BibitemShut
  {NoStop}%
\bibitem [{\citenamefont {Chu}\ \emph {et~al.}(2018)\citenamefont {Chu},
  \citenamefont {Kharel}, \citenamefont {Yoon}, \citenamefont {Frunzio},
  \citenamefont {Rakich},\ and\ \citenamefont {Schoelkopf}}]{chu2018}%
  \BibitemOpen
  \bibfield  {author} {\bibinfo {author} {\bibfnamefont {Y.}~\bibnamefont
  {Chu}}, \bibinfo {author} {\bibfnamefont {P.}~\bibnamefont {Kharel}},
  \bibinfo {author} {\bibfnamefont {T.}~\bibnamefont {Yoon}}, \bibinfo {author}
  {\bibfnamefont {L.}~\bibnamefont {Frunzio}}, \bibinfo {author} {\bibfnamefont
  {P.~T.}\ \bibnamefont {Rakich}},\ and\ \bibinfo {author} {\bibfnamefont
  {R.~J.}\ \bibnamefont {Schoelkopf}},\ }\bibfield  {title} {\bibinfo {title}
  {Creation and control of multi-phonon {{Fock}} states in a bulk acoustic-wave
  resonator},\ }\href {https://doi.org/10.1038/s41586-018-0717-7} {\bibfield
  {journal} {\bibinfo  {journal} {Nature}\ }\textbf {\bibinfo {volume} {563}},\
  \bibinfo {pages} {666} (\bibinfo {year} {2018})}\BibitemShut {NoStop}%
\bibitem [{\citenamefont {Wollack}\ \emph {et~al.}(2022)\citenamefont
  {Wollack}, \citenamefont {Cleland}, \citenamefont {Gruenke}, \citenamefont
  {Wang}, \citenamefont {{Arrangoiz-Arriola}},\ and\ \citenamefont
  {{Safavi-Naeini}}}]{wollack2022}%
  \BibitemOpen
  \bibfield  {author} {\bibinfo {author} {\bibfnamefont {E.~A.}\ \bibnamefont
  {Wollack}}, \bibinfo {author} {\bibfnamefont {A.~Y.}\ \bibnamefont
  {Cleland}}, \bibinfo {author} {\bibfnamefont {R.~G.}\ \bibnamefont
  {Gruenke}}, \bibinfo {author} {\bibfnamefont {Z.}~\bibnamefont {Wang}},
  \bibinfo {author} {\bibfnamefont {P.}~\bibnamefont {{Arrangoiz-Arriola}}},\
  and\ \bibinfo {author} {\bibfnamefont {A.~H.}\ \bibnamefont
  {{Safavi-Naeini}}},\ }\bibfield  {title} {\bibinfo {title} {Quantum state
  preparation and tomography of entangled mechanical resonators},\ }\href
  {https://doi.org/10.1038/s41586-022-04500-y} {\bibfield  {journal} {\bibinfo
  {journal} {Nature}\ }\textbf {\bibinfo {volume} {604}},\ \bibinfo {pages}
  {463} (\bibinfo {year} {2022})}\BibitemShut {NoStop}%
\bibitem [{\citenamefont {Schrinski}\ \emph {et~al.}(2023)\citenamefont
  {Schrinski}, \citenamefont {Yang}, \citenamefont {{von L{\"u}pke}},
  \citenamefont {Bild}, \citenamefont {Chu}, \citenamefont {Hornberger},
  \citenamefont {Nimmrichter},\ and\ \citenamefont {Fadel}}]{schrinski2023}%
  \BibitemOpen
  \bibfield  {author} {\bibinfo {author} {\bibfnamefont {B.}~\bibnamefont
  {Schrinski}}, \bibinfo {author} {\bibfnamefont {Y.}~\bibnamefont {Yang}},
  \bibinfo {author} {\bibfnamefont {U.}~\bibnamefont {{von L{\"u}pke}}},
  \bibinfo {author} {\bibfnamefont {M.}~\bibnamefont {Bild}}, \bibinfo {author}
  {\bibfnamefont {Y.}~\bibnamefont {Chu}}, \bibinfo {author} {\bibfnamefont
  {K.}~\bibnamefont {Hornberger}}, \bibinfo {author} {\bibfnamefont
  {S.}~\bibnamefont {Nimmrichter}},\ and\ \bibinfo {author} {\bibfnamefont
  {M.}~\bibnamefont {Fadel}},\ }\bibfield  {title} {\bibinfo {title}
  {Macroscopic {{Quantum Test}} with {{Bulk Acoustic Wave Resonators}}},\
  }\href {https://doi.org/10.1103/PhysRevLett.130.133604} {\bibfield  {journal}
  {\bibinfo  {journal} {Physical Review Letters}\ }\textbf {\bibinfo {volume}
  {130}},\ \bibinfo {pages} {133604} (\bibinfo {year} {2023})}\BibitemShut
  {NoStop}%
\bibitem [{\citenamefont {Bild}\ \emph {et~al.}(2023)\citenamefont {Bild},
  \citenamefont {Fadel}, \citenamefont {Yang}, \citenamefont {{von L{\"u}pke}},
  \citenamefont {Martin}, \citenamefont {Bruno},\ and\ \citenamefont
  {Chu}}]{bild2023}%
  \BibitemOpen
  \bibfield  {author} {\bibinfo {author} {\bibfnamefont {M.}~\bibnamefont
  {Bild}}, \bibinfo {author} {\bibfnamefont {M.}~\bibnamefont {Fadel}},
  \bibinfo {author} {\bibfnamefont {Y.}~\bibnamefont {Yang}}, \bibinfo {author}
  {\bibfnamefont {U.}~\bibnamefont {{von L{\"u}pke}}}, \bibinfo {author}
  {\bibfnamefont {P.}~\bibnamefont {Martin}}, \bibinfo {author} {\bibfnamefont
  {A.}~\bibnamefont {Bruno}},\ and\ \bibinfo {author} {\bibfnamefont
  {Y.}~\bibnamefont {Chu}},\ }\bibfield  {title} {\bibinfo {title}
  {Schr{\"o}dinger cat states of a 16-microgram mechanical oscillator},\ }\href
  {https://doi.org/10.1126/science.adf7553} {\bibfield  {journal} {\bibinfo
  {journal} {Science}\ }\textbf {\bibinfo {volume} {380}},\ \bibinfo {pages}
  {274} (\bibinfo {year} {2023})}\BibitemShut {NoStop}%
\bibitem [{\citenamefont {Haroche}\ and\ \citenamefont
  {Raimond}(2006)}]{haroche2006}%
  \BibitemOpen
  \bibfield  {author} {\bibinfo {author} {\bibfnamefont {S.}~\bibnamefont
  {Haroche}}\ and\ \bibinfo {author} {\bibfnamefont {J.-M.}\ \bibnamefont
  {Raimond}},\ }\href@noop {} {\emph {\bibinfo {title} {Exploring the
  {{Quantum}}: {{Atoms}}, {{Cavities}}, and {{Photons}}}}}\ (\bibinfo
  {publisher} {OUP Oxford},\ \bibinfo {year} {2006})\BibitemShut {NoStop}%
\bibitem [{\citenamefont {Liao}\ and\ \citenamefont {Nori}(2014)}]{liao2014}%
  \BibitemOpen
  \bibfield  {author} {\bibinfo {author} {\bibfnamefont {J.-Q.}\ \bibnamefont
  {Liao}}\ and\ \bibinfo {author} {\bibfnamefont {F.}~\bibnamefont {Nori}},\
  }\bibfield  {title} {\bibinfo {title} {Spectrometric reconstruction of
  mechanical-motional states in optomechanics},\ }\href
  {https://doi.org/10.1103/PhysRevA.90.023851} {\bibfield  {journal} {\bibinfo
  {journal} {Physical Review A}\ }\textbf {\bibinfo {volume} {90}},\ \bibinfo
  {pages} {023851} (\bibinfo {year} {2014})}\BibitemShut {NoStop}%
\end{thebibliography}%


\begin{thebibliography}{2}%
\makeatletter
\providecommand \@ifxundefined [1]{%
 \@ifx{#1\undefined}
}%
\providecommand \@ifnum [1]{%
 \ifnum #1\expandafter \@firstoftwo
 \else \expandafter \@secondoftwo
 \fi
}%
\providecommand \@ifx [1]{%
 \ifx #1\expandafter \@firstoftwo
 \else \expandafter \@secondoftwo
 \fi
}%
\providecommand \natexlab [1]{#1}%
\providecommand \enquote  [1]{``#1''}%
\providecommand \bibnamefont  [1]{#1}%
\providecommand \bibfnamefont [1]{#1}%
\providecommand \citenamefont [1]{#1}%
\providecommand \href@noop [0]{\@secondoftwo}%
\providecommand \href [0]{\begingroup \@sanitize@url \@href}%
\providecommand \@href[1]{\@@startlink{#1}\@@href}%
\providecommand \@@href[1]{\endgroup#1\@@endlink}%
\providecommand \@sanitize@url [0]{\catcode `\\12\catcode `\$12\catcode
  `\&12\catcode `\#12\catcode `\^12\catcode `\_12\catcode `\%12\relax}%
\providecommand \@@startlink[1]{}%
\providecommand \@@endlink[0]{}%
\providecommand \url  [0]{\begingroup\@sanitize@url \@url }%
\providecommand \@url [1]{\endgroup\@href {#1}{\urlprefix }}%
\providecommand \urlprefix  [0]{URL }%
\providecommand \Eprint [0]{\href }%
\providecommand \doibase [0]{https://doi.org/}%
\providecommand \selectlanguage [0]{\@gobble}%
\providecommand \bibinfo  [0]{\@secondoftwo}%
\providecommand \bibfield  [0]{\@secondoftwo}%
\providecommand \translation [1]{[#1]}%
\providecommand \BibitemOpen [0]{}%
\providecommand \bibitemStop [0]{}%
\providecommand \bibitemNoStop [0]{.\EOS\space}%
\providecommand \EOS [0]{\spacefactor3000\relax}%
\providecommand \BibitemShut  [1]{\csname bibitem#1\endcsname}%
\let\auto@bib@innerbib\@empty
\bibitem [{\citenamefont {{de Oliveira}}\ \emph {et~al.}(1990)\citenamefont
  {{de Oliveira}}, \citenamefont {Kim}, \citenamefont {Knight},\ and\
  \citenamefont {Buek}}]{deoliveira1990}%
  \BibitemOpen
  \bibfield  {author} {\bibinfo {author} {\bibfnamefont {F.~A.~M.}\
  \bibnamefont {{de Oliveira}}}, \bibinfo {author} {\bibfnamefont {M.~S.}\
  \bibnamefont {Kim}}, \bibinfo {author} {\bibfnamefont {P.~L.}\ \bibnamefont
  {Knight}},\ and\ \bibinfo {author} {\bibfnamefont {V.}~\bibnamefont {Buek}},\
  }\bibfield  {title} {\bibinfo {title} {Properties of displaced number
  states},\ }\href {https://doi.org/10.1103/PhysRevA.41.2645} {\bibfield
  {journal} {\bibinfo  {journal} {Physical Review A}\ }\textbf {\bibinfo
  {volume} {41}},\ \bibinfo {pages} {2645} (\bibinfo {year}
  {1990})}\BibitemShut {NoStop}%
\bibitem [{\citenamefont {Wunsche}(1991)}]{wunsche1991}%
  \BibitemOpen
  \bibfield  {author} {\bibinfo {author} {\bibfnamefont {A.}~\bibnamefont
  {Wunsche}},\ }\bibfield  {title} {\bibinfo {title} {Displaced {{Fock}} states
  and their connection to quasiprobabilities},\ }\href
  {https://doi.org/10.1088/0954-8998/3/6/005} {\bibfield  {journal} {\bibinfo
  {journal} {Quantum Optics: Journal of the European Optical Society Part B}\
  }\textbf {\bibinfo {volume} {3}},\ \bibinfo {pages} {359} (\bibinfo {year}
  {1991})}\BibitemShut {NoStop}%
\end{thebibliography}%

\end{document}



\title{Supplemental Material: Nonclassical mechanical states in cavity optomechanics in the single-photon strong-coupling regime}

\author{Jonathan L. Wise}
\email{jonathan.wise@u-bordeaux.fr}
\author{Clément Dutreix}%
\author{Fabio Pistolesi}%

\affiliation{%
 Université de Bordeaux, CNRS, LOMA, UMR 5798, F-33400 Talence, France
}%




\date{\today}

\maketitle

\onecolumngrid
\appendix
\renewcommand\thefigure{\thesection.\arabic{figure}}


%
We repeat here for clarity the driven optomechanical hamiltonian in the lab frame rotating at the laser drive frequency:
\begin{equation}
\hat{H} = -\Delta\hat{a}^\dagger\hat{a} + \Omega_m\hat{b}^\dagger\hat{b} + g_0\hat{a}^\dagger\hat{a}\left(\hat{b}+\hat{b}^\dagger\right) + \varepsilon \left( \hat{a}+ \hat{a}^\dagger \right),
\label{eq:H}
\end{equation}
where $\hat{a}$ and $\hat{b}$ are the annihilation operators for the cavity field and oscillator mode, respectively, $\Delta=\omega_L - \omega_c$ is the laser detuning from the cavity frequency, $\Omega_m$ is the mechanical mode frequency, $g_0$ is the single photon optomechanical coupling strength and $\varepsilon$ is proportional to the intensity of the laser drive. 
In the Lang-Firsov frame, achieved by the unitary transformation $\hat{H}_\mathrm{LF} = \hat{U}_\mathrm{LF}\hat{H}\hat{U}_\mathrm{LF}^\dagger$, with $\hat{U}_\mathrm{LF} = e^{\tilde{g}_0 \hat{a}^\dagger\hat{a}(\hat{b}^\dagger-\hat{b})}$ and $\tilde{g}_0=g_0/\Omega_m$, the transformed hamiltonian reads
\begin{subequations}\label{eq:H_LF}
\begin{align}
\hat{H}_\mathrm{LF} &= \hat{H}_0 + \hat{V}, \tag{\ref{eq:H_LF}} \\ 
\hat{H}_0 &= -\Delta \hat{a}^\dagger\hat{a} - \mathcal{K}\left( \hat{a}^\dagger\hat{a}\right)^2 +\Omega_m\hat{b}^\dagger\hat{b}, \label{eq:H_0} \\
\hat{V} &= \varepsilon \left[\hat{D}(\tilde{g}_0)\hat{a}^\dagger + \hat{D}(-\tilde{g}_0)\hat{a}\right], \label{eq:V}
\end{align}
\end{subequations}
where $\mathcal{K}=g_0^2/\Omega_m$ is the Kerr nonlinearity strength, and $\hat{D}(\beta) = e^{\beta \hat{b}^\dagger - \beta^* \hat{b}}$ is the mechanical displacement operator. 

\section{Undriven case  -- periodicity}
\setcounter{figure}{0}
For the undriven system ($\varepsilon=0$), we show in Eq.~(4) of the main text that the dynamics of the oscillator are periodic with period $2\pi/\Omega_m$. 
In the main text this is shown for the specific case of initial coherent states for both cavity and oscillator, but this result may be easily generalised to any initial state, as we show here. 
We take the general initial state in the lab frame
\begin{equation}
\vert \xi(t=0)\rangle = \sum_{n,m=0}^\infty A_{nm}\vert n\rangle_a \otimes \vert m\rangle_b,
\end{equation}
where the summation indices $n$ and $m$ run over all the Fock states of the cavity and oscillator systems, respectively. 
We compute the subsequent evolution by transforming to the Lang-Firsov basis, applying the diagonal time evolution and transforming back, i.e. $\vert \xi(t)\rangle = \hat{U}_\mathrm{LF}^{-1}e^{-i\hat{H}_0 t}\hat{U}_\mathrm{LF} \vert \xi(t=0)\rangle$, resulting in a general expression for the wavefunction
\begin{equation}
\vert \xi(t) \rangle = \sum_{n,m=0}^\infty \tilde{A}_{nm} e^{iE_n^{(a)}t} \vert n \rangle_a \otimes \sum_{p=0}^\infty B_{pmn} e^{-i\Omega_m m t} \vert p \rangle_b,  
\end{equation}
where the nonlinear cavity energy is $E_n^{(a)} = \Delta n + \mathcal{K}n^2$. 
The effect of transforming to and from the Lang-Firsov has been captured by introducing the shorthand:
\begin{align}
\tilde{A}_{nm} &= \sum_{q=0}^\infty A_{nq} \langle m \vert_b e^{\tilde{g}_0 n (\hat{b}^\dagger - \hat{b})}\vert q \rangle_b,  \\  
B_{pmn} &= \langle p \vert_b e^{-\tilde{g}_0 n (\hat{b}^\dagger - \hat{b})} \vert m\rangle_b.
\end{align}
Constructing the system density matrix as $\rho = \vert \xi \rangle \langle \xi \vert$ and tracing over the cavity field gives for the oscillator
\begin{align}
\rho_b(t) &= \sum_{n=0}^\infty \langle n \vert_a \rho \vert n \rangle_a \nonumber \\ 
&= \sum_{n=0}^\infty \sum_{m, m'=0}^\infty \sum_{p, p'=0}^\infty \tilde{A}_{nm}\tilde{A}_{nm'}^* B_{pmn}B_{p'm'n}^* e^{-i\Omega_m(m-m')t}\vert p \rangle_b \langle p' \vert_b,
\end{align}
which is clearly periodic in $2\pi/\Omega_m$. 
This result is what motivates the final step in our nonclassical state generation method: once the system has been driven to a sufficiently nonclassical state, we propose switching off the drive.
If dissipation effects are negligible then this leads to a dynamics where the nonclassical state reemerges every period, $2\pi/\Omega_m$. 

\section{Driven case -- weak drive perturbative solution}
\setcounter{figure}{0}
For the driven system ($\varepsilon\ne 0$) we apply perturbation theory in the interaction picture defined by Eq.~(\ref{eq:H_LF}) in the Lang-Firsov basis. We have the Schrödinger equation 
\begin{equation}
i\partial_t \vert\tilde{\psi}(t)\rangle_\mathrm{LF} = \hat{\tilde{V}}(t) \vert\tilde{\psi}(t)\rangle_\mathrm{LF},
\end{equation}
where $\vert\tilde{\psi}(t)\rangle_\mathrm{LF} = e^{i\hat{H}_0t}\hat{U}_\mathrm{LF}\vert\psi(t)\rangle$ and $\hat{\tilde{V}}(t) = e^{i\hat{H}_0t}\hat{V}e^{-i\hat{H}_0t}$ are the interaction picture wavefunction and perturbation operator, respectively. 
Note that in the main text the subscript $\mathrm{LF}$ is omitted.  
Up to first order in $\hat{\tilde{V}}$ the time evolution operator defined by the expression $\vert\tilde{\psi}(t)\rangle_\mathrm{LF}= \hat{U}(t) \vert\psi(0)\rangle_\mathrm{LF}$ is given by
\begin{equation}
\hat{U}(t) \approx 1-i\int_0^t d\tau \hat{\tilde{V}}(\tau).
\end{equation}
%
An expression for the wavefunction in the Schrödinger picture in the Lang Firsov basis may then be calculated as:
\begin{align}
\vert \psi(t)\rangle_\mathrm{LF} &= e^{-i\hat{H}_0t}\hat{U}(t)\vert\psi(0)\rangle_\mathrm{LF} \nonumber \\
&\approx \mathcal{N}\left[ \vert\psi_0(t)\rangle_\mathrm{LF} - i\tilde{\varepsilon}  \vert\psi_1(t)\rangle_\mathrm{LF}\right],
\end{align}
where $\mathcal{N}$ is a normalization factor and $\tilde{\varepsilon} = \varepsilon/\Omega_m$. We suppose an initial state in the (rotating) lab frame $\vert\psi(t=0)\rangle=\sum_{n} a_n \vert n\rangle_a \otimes \vert \beta\rangle_b$, where for an initial cavity coherent state $\vert \alpha \rangle_a$ we have $a_n=e^{-|\alpha|^2} \alpha^n/\sqrt{n!}$. 
The zeroth order term is then merely the undriven result, given in the Lang Firsov basis as:
\begin{align}
\vert\psi_0(t)\rangle_\mathrm{LF} &= e^{-i\hat{H}_0t}\vert \psi(0)\rangle_\mathrm{LF} \nonumber \\ 
&=   e^{-i\hat{H}_0t} \sum_{n=0}^\infty A_n \vert n\rangle_a \otimes \vert\beta + n\tilde{g}_0\rangle_b \nonumber \\ 
&= \sum_{n=0}^\infty A_n e^{i(\Delta n + \mathcal{K}n^2)t}\vert n\rangle_a \otimes \vert \bar{\beta}_n e^{-i\Omega_m t}\rangle_b,
\end{align}
where the factor $A_n=a_n e^{-i\tilde{g}_0n \mathrm{Im} \beta}$ and $\bar{\beta}_n = \beta + n \tilde{g}_0$. 
Returning to the lab frame yields the expression
\begin{equation}
\vert \psi_0(t) \rangle = \sum_{n=0}^\infty a_n e^{i\phi_n(t)} \vert n\rangle_a \otimes \vert \beta_n(t) \rangle_b,
\end{equation}
which corresponds to Eq.~(3) of the main text, and we specify here the phase omitted from the main text:
\begin{equation}
\phi_n(t) = \Delta n t + \mathcal{K} n^2\left(t -\frac{\sin\Omega_m t}{\Omega_m} \right) + \tilde{g}_0 n \mathrm{Im} \left[ \beta(e^{-i\Omega_m t}-1)\right].
\end{equation}
The first order correction is given by the expression:
\begin{equation}
\vert\psi_1(t)\rangle_\mathrm{LF} = e^{-i\hat{H}_0t} \int_0^{\Omega_m t} d\tilde{\tau}\, e^{i\hat{H}_0\tau} \left[D(\tilde{g}_0)\hat{a}^\dagger + D(-\tilde{g}_0)\hat{a}\right] e^{-i\hat{H}_0\tau} \vert\psi(0)\rangle_\mathrm{LF},
\end{equation}
where $\tilde{\tau} = \Omega_m \tau$ and the integrand is found to be given by:
\begin{align}
&e^{i\hat{H}_0\tau} \left[D(\tilde{g}_0)\hat{a}^\dagger + D(-\tilde{g}_0)\hat{a}\right] e^{-i\hat{H}_0\tau} \vert\psi(0)\rangle_\mathrm{LF} \nonumber \\ 
&= \sum_{n=0}^\infty A_n \left[e^{i \varphi_n^-(\tau)}\sqrt{n} \vert n-1\rangle_a \otimes \vert \bar{\beta}_n - \tilde{g}_0 e^{i\Omega_m \tau}\rangle_b + e^{i \varphi_n^+(\tau)}\sqrt{n+1} \vert n+1\rangle_a  \otimes \vert \bar{\beta}_n + \tilde{g}_0 e^{i\Omega_m \tau}\rangle_b \right],
\end{align}
where the phases are given by:
\begin{equation}
\varphi_n^\pm(\tau) = \mp \Delta\tau \mp \mathcal{K}(2n\pm1)\tau \pm \tilde{g}_0\mathrm{Im} \left[\bar{\beta}_n^*e^{i\Omega_m \tau}\right].
\end{equation}
This is the explicit expression for the phases appearing in Eq.~(5) of the main text. 
Relabelling the dummy indices in the cavity state sums, we may write the total solution for the wavefunction in the interaction picture of the Lang Firsov basis [c.f. Eq.~(5) of the main text]:
\begin{align}
\vert \tilde{\psi}(t)\rangle_\mathrm{LF} \approx \mathcal{N}\sum_{n=0}^\infty &\vert n\rangle_a \otimes \left[A_n \vert \bar{\beta}_n\rangle_b -i \tilde{\varepsilon} \int_0^{\Omega_m t} d\tilde{\tau}\, \left( \sqrt{n}A_{n-1} e^{i\varphi_{n-1}^+(\tau)} \vert \bar{\beta}_{n-1}^+(\tau)\rangle_b + \sqrt{n+1}A_{n+1} e^{i\varphi_{n+1}^-(\tau)} \vert\bar{\beta}_{n+1}^-(\tau)\rangle_b  \right) \right],
\label{eq:psi_LF}
\end{align}
where we denoted the time-dependent coherent states $\bar{\beta}_n^\pm(\tau) = \bar{\beta}_n \pm \tilde{g}_0e^{i\Omega_m\tau}$. 
By expanding the exponentials as power series one can perform the integrals analytically, however this leads to a rather heavy result given in terms of multiple infinite sums.
Here, we include this solution but first we compute the approximate short time expression shown in Eq.~(8) of the main text.

\subsection{Short time expression in the interaction picture, Lang-Firsov basis}
For $\Omega_m t \ll 1$ the coherent state parameters are expanded as $\bar{\beta}_{n\mp1}^\pm \approx \bar{\beta}_n + \gamma_{\pm}$, where $\gamma_\pm = \pm i \tilde{g}_0 \tilde{\tau}$ and $\tilde{\tau} = \Omega_m \tau$. 
Then, assuming $|\gamma_\pm| \ll |\bar{\beta}_n|$ the coherent states in the integrand may be expanded via two-dimensional Taylor expansion:
\begin{align}
\vert\bar{\beta}_n + \gamma_\pm\rangle \approx \vert\bar{\beta}_n\rangle + \gamma_\pm \partial_\alpha \vert\alpha\rangle|_{\alpha = \bar{\beta}_n} + \gamma_\pm^* \partial_{\alpha^*} \vert\alpha\rangle |_{\alpha^* = \bar{\beta}_n^*},
\end{align}
where to compute the derivatives of the coherent state one may use the Baker-Campbell-Hausdorff formula to write: $\vert\alpha\rangle = e^{\alpha\hat{b}^\dagger - \alpha^* \hat{b}}\vert 0\rangle = e^{-\alpha \alpha^*/2}e^{\alpha\hat{b}^\dagger}e^{-\alpha^*\hat{b}}\vert0 \rangle$ or alternatively $\vert\alpha \rangle = e^{\alpha \alpha^*/2}e^{-\alpha^*\hat{b}}e^{\alpha\hat{b}^\dagger}\vert 0\rangle$, and hence find
\begin{align}
\partial_\alpha \vert\alpha\rangle &= \left( -\alpha^*/2 + \hat{b}^\dagger \right)\vert \alpha\rangle, \\ 
\partial_{\alpha^*} \vert\alpha\rangle &= \left( \alpha/2 - \hat{b} \right)\vert\alpha\rangle.
\end{align}
Expanding also the exponential phase factors in the integrand for $\Omega_m t \ll 1$ and collecting powers of the rescaled integration variable $\tilde{\tau}$, leads to the expression:
\begin{align}
\vert\tilde{\psi}(t)\rangle_\mathrm{LF} \approx \mathcal{N}\sum_{n=0}^\infty  \vert n\rangle_a &\otimes \left[A_n \vert\bar{\beta}_n \rangle_b
-\tilde{\varepsilon} e^{-i\tilde{g}_0 n \mathrm{Im}\beta} \left\{ \sqrt{n}\int_0^{\Omega_m t} d\tilde{\tau} \left[ 1 - i \left(\tilde{\Delta} + 2n\tilde{\mathcal{K}} - i \tilde{g}_0(\hat{b}^\dagger + \hat{b})\right) \tilde{\tau} + \mathcal{O}(\tilde{\tau}^2) \right] \right. \right. \nonumber \\
& \left. \left. + \sqrt{n+1}\int_0^{\Omega_m t} d\tilde{\tau} \left[ 1 + i \left(\tilde{\Delta} + 2n\tilde{\mathcal{K}} - i \tilde{g}_0(\hat{b}^\dagger + \hat{b})\right) \tilde{\tau} + \mathcal{O}(\tilde{\tau}^2) \right]   \right\}\vert\bar{\beta}_n\rangle_b \right],
\end{align}
where frequencies with tildes are adimensionalised by $\Omega_m$. 
Performing the integration yields the result
\begin{equation}
\vert\tilde{\psi}(t)\rangle_\mathrm{LF} \approx \mathcal{N} \sum_{n=0}^\infty \vert n\rangle_a\otimes \left[\tilde{A}_n \vert\bar{\beta}_n\rangle_b + \tilde{B}_n\left(\hat{b}^\dagger +\hat{b}  \right)\vert \bar{\beta}_n\rangle_b \right],
\label{eq:psi_short_op}
\end{equation}
where the coefficients are
\begin{align}
\tilde{A}_n &= A_n - i\tilde{\varepsilon}e^{-i\tilde{g}_0 n \mathrm{Im}\beta } \Omega_m t\left[ \sqrt{n}a_{n-1}\left(1-i(\tilde{\Delta}+2n\tilde{\mathcal{K}})\Omega_m t/2\right) + \sqrt{n+1}a_{n+1}\left(1+i(\tilde{\Delta}+2n\tilde{\mathcal{K}})\Omega_m t/2\right) \right], \\
\tilde{B}_n &= \tilde{\varepsilon}\tilde{g}_0 e^{-i\tilde{g}_0 n \mathrm{Im}\beta } \frac{(\Omega_m t)^2}{2} \left[ \sqrt{n}a_{n-1} -\sqrt{n+1}a_{n+1}\right].
\end{align}
To arrive at the expression (8) of the main text we make a further approximation in order to re-express the unusual $\hat{b}^\dagger \vert \bar{\beta}_n\rangle_b$ term. 
Specifically, we notice that by definition $\bar{\beta}_{n\pm1} = \bar{\beta}_n \pm \tilde{g}_0$, and for $|\tilde{g}_0/\bar{\beta}_n|\ll 1$ we may expand:
\begin{align}
\vert\bar{\beta}_{n\pm1}\rangle_b &= \chi_{\pm}e^{-|\bar{\beta}_n|^2/2}\sum_{m=0}^\infty \bar{\beta}_n^m \left(1\pm \frac{\tilde{g}_0}{\bar{\beta}_n}\right)^m \frac{\left.\hat{b}^\dagger\right.^m}{m!}\vert 0\rangle \nonumber \\
&\approx \chi_\pm e^{-|\bar{\beta}_n|^2/2} \sum_{m=0}^\infty \bar{\beta}_n^m\left(1\pm m \frac{\tilde{g}_0}{\bar{\beta}_n}\right)\frac{\left.\hat{b}^\dagger\right.^m}{m!}\vert 0\rangle \nonumber \\ 
&= \chi_\pm \left[\vert\bar{\beta}_n\rangle_b \pm \tilde{g}_0\hat{b}^\dagger\vert \bar{\beta}_n\rangle_b\right], 
\label{eq:beta_npm1}
\end{align}
where we defined $\chi_\pm = e^{-\tilde{g}_0^2/2 \mp \tilde{g}_0\mathrm{Re}\bar{\beta}_n}$.
In the expansion we require $|\tilde{g}_0/\bar{\beta}_n|\ll 1$, which can be achieved for all relevant $n$ simply by choosing a sufficiently large initial coherent state $|\beta|\gg \tilde{g}_0$ for the oscillator. 
We may invert Eq.~(\ref{eq:beta_npm1}) and write the sum:
\begin{equation}
\vert \bar{\beta}_{n+1}\rangle_b + \vert \bar{\beta}_{n-1}\rangle_b \approx \left(\chi_+ + \chi_-\right)\vert \bar{\beta}_n\rangle_b + \tilde{g}_0\left(\chi_+ - \chi_-\right) \hat{b}^\dagger \vert\bar{\beta}_n\rangle_b.
\end{equation}
Substituting for $\hat{b}^\dagger\vert\bar{\beta}_n\rangle_b$ from the above into the expression for the wavefunction Eq.~\eqref{eq:psi_short_op} gives
\begin{equation}
\vert \tilde{\psi}(t)\rangle_\mathrm{LF} \approx \mathcal{N} \sum_{n=0}^\infty \vert n\rangle_a \otimes \left[B_n^0 \vert \bar{\beta}_n\rangle_b  + B_n^+ \vert \bar{\beta}_{n+1}\rangle_b + B_n^-\vert\bar{\beta}_{n-1}\rangle_b \right],
\label{eq:psi_short_SP}
\end{equation}
as in Eq.~(8) of the main text, with the coefficients:
\begin{align}
B_n^0 &= \tilde{A}_n+\tilde{B}_n\left[\beta_n-\frac{\chi_++\chi_-}{\tilde{g}_0\left(\chi_+-\chi_-\right)}\right],\\ 
B_n^+ &= \frac{\tilde{B}_n}{\tilde{g}_0\left(\chi_+-\chi_-\right)}, \\ 
B_n^-&=B_n^+.
\end{align} 

\subsection{Full solution in the Schrödinger picture, lab basis}
Returning to the full solution Eq.~(\ref{eq:psi_LF}), it is convenient to first return to the Schrödinger picture and lab basis before performing the integration. 
Going back to the Schrödinger picture and lab frame just introduces additional phase factors and a shift for the mechanical coherent states:
\begin{align}
\vert\psi(t)\rangle &= e^{-\tilde{g}_0\hat{a}^\dagger\hat{a}(\hat{b}^\dagger-\hat{b})}e^{-i\hat{H}_0 t}\vert\tilde{\psi}(t)\rangle_\mathrm{LF} \\ 
= \mathcal{N}& \sum_{n=0}^\infty e^{i(\Delta n + \mathcal{K}n^2)t} \vert n\rangle_a \otimes \left[A_n e^{-i\tilde{g}_0 n \mathrm{Im}(\bar{\beta}_n^*e^{i\Omega_m t})}\vert\bar{\beta}_n e^{-i\Omega_m t}-\tilde{g}_0 n\rangle_b \phantom{\int_0^t} \right. \nonumber \\
&\left. -i \tilde{\varepsilon} \int_0^{\Omega_m t} d\tilde{\tau}\, \left( \sqrt{n}A_{n-1}e^{i\varphi_{n-1}^+(\tau)} e^{-i \tilde{g}_0 n \mathrm{Im}[\bar{\beta}_{n-1}^+(\tau)^*e^{i\Omega_m t}]}\vert\bar{\beta}_{n-1}^+(\tau)e^{-i\Omega_m t}-\tilde{g}_0 n\rangle_b \right. \right. \nonumber \\ 
& \left. \left. + \sqrt{n+1}A_{n+1}e^{i\varphi_{n+1}^-(\tau)}e^{-i\tilde{g}_0 n\mathrm{Im}[\bar{\beta}_{n+1}^-(\tau)^*e^{i\Omega_m t}]}  \vert\bar{\beta}_{n+1}^-(\tau)e^{-i\Omega_m t}- \tilde{g}_0 n\rangle_b  \right) \right].
\label{eq:psi_lab}
\end{align}
In order to perform the integrals, we can express the $\tau$-dependence of the coherent states as phase factors, giving for the overall result
\begin{align}
\vert\psi(t)\rangle & = \mathcal{N}\sum_{n=0}^\infty e^{i(\Delta n + \mathcal{K}n^2)t} \vert n\rangle_a \otimes \left\{ A_n e^{-i\tilde{g}_0 n \mathrm{Im}(\bar{\beta}_n^*e^{i\Omega_m t})}\vert\bar{\beta}_n e^{-i\Omega_m t}\rangle_b \phantom{\sum_{k=0}^m} \right. && \nonumber \\
&\left. \phantom{\sum_{k=0}^m}  -i \tilde{\varepsilon} \sum_{m=0}^\infty \frac{\vert m\rangle_b}{\sqrt{m!}}\sum_{k=0}^m 
\begin{pmatrix}
m \\
k
\end{pmatrix}
e^{-i\Omega_m k t}  
\left[ \sqrt{n} A_{n-1} e^{\xi_n^+(t)} 
\left( \bar{\beta}_{n-1}e^{-i\Omega_m t}-\tilde{g}_0 n\right)^{m-k}(+\tilde{g}_0)^k \mathcal{I}_{nk}^+\right. \right. \nonumber \\ 
& \left. \phantom{\sum_{k=0}^m} \left. \hspace{5cm} + \sqrt{n+1} A_{n+1} e^{\xi_n^-(t)} 
\left( \bar{\beta}_{n+1}e^{-i\Omega_m t}-\tilde{g}_0 n\right)^{m-k}(-\tilde{g}_0)^k \mathcal{I}_{nk}^- 
\right] \right\}  ,
\label{eq:psi_lab_explicit}
\end{align}
where we introduced the shorthand
\begin{equation}
\xi_n^\pm(t) = -i\tilde{g}_0 n \mathrm{Im}[\bar{\beta}_{n\mp1}^*e^{i\Omega_m t}] -\frac{1}{2}\left(|\bar{\beta}_{n\mp1}|^2 + \tilde{g}_0^2(n^2+1) - 2\tilde{g}_0 n \mathrm{Re}[\bar{\beta}_{n\mp1}e^{-i\Omega_m t}]\right), 
\end{equation}
and the integrals to be performed are
\begin{equation}
\mathcal{I}_{nk}^\pm = \int_0^{\Omega_m t} d\tilde{\tau} e^{iC_{nk}^\pm \tilde{\tau} + Q_n^\pm e^{i\tilde{\tau}} + P_n^\pm e^{-i\tilde{\tau}}},
\end{equation}
with the coefficients
\begin{align}
C_{nk}^\pm &= \mp \tilde{\Delta} \mp \tilde{\mathcal{K}}(2n\mp 1) + k, \\ 
Q_n^\pm &= \pm \tilde{g}_0^2 n e^{-i\Omega_m t}, \\ 
P_n^\pm &= \mp \tilde{g}_0 \bar{\beta}_{n\mp1}.
\end{align}
These integrals may be performed by expanding the exponentials as infinite power series, ultimately leading to an explicit expression for the wavefunction. 
However, in practice we found it more numerically efficient to use Eq.~(\ref{eq:psi_lab}) and approximate the integrals as sums over a discretised time interval --  care was taken to ensure the discretisation was sufficiently fine to match the exact result of the integral. 

\section{Comparison of analytics to numerics}
\setcounter{figure}{0} 
We note that the numerical results were computed using QuTiP with a truncated Hilbert space of size $N_c\times N_m$. 
For the dissipationless case this meant solving the Schrödinger equation for the full system wavefunction, and including dissipation meant solving the master equation for the system density matrix. 
To avoid effects related to the truncation of the Hilbert space we computed the average value of the commutators $\langle[ \hat{a}, \hat{a}^\dagger] \rangle$ and $\langle [ \hat{b}, \hat{b}^\dagger] \rangle$. 
In practice, ensuring that the commutators remained close to 1 for the parameter ranges explored meant taking $N_c = 8-12$ and $N_m=300-400$.

\begin{figure}
\centering
\includegraphics[width = 0.9\textwidth]{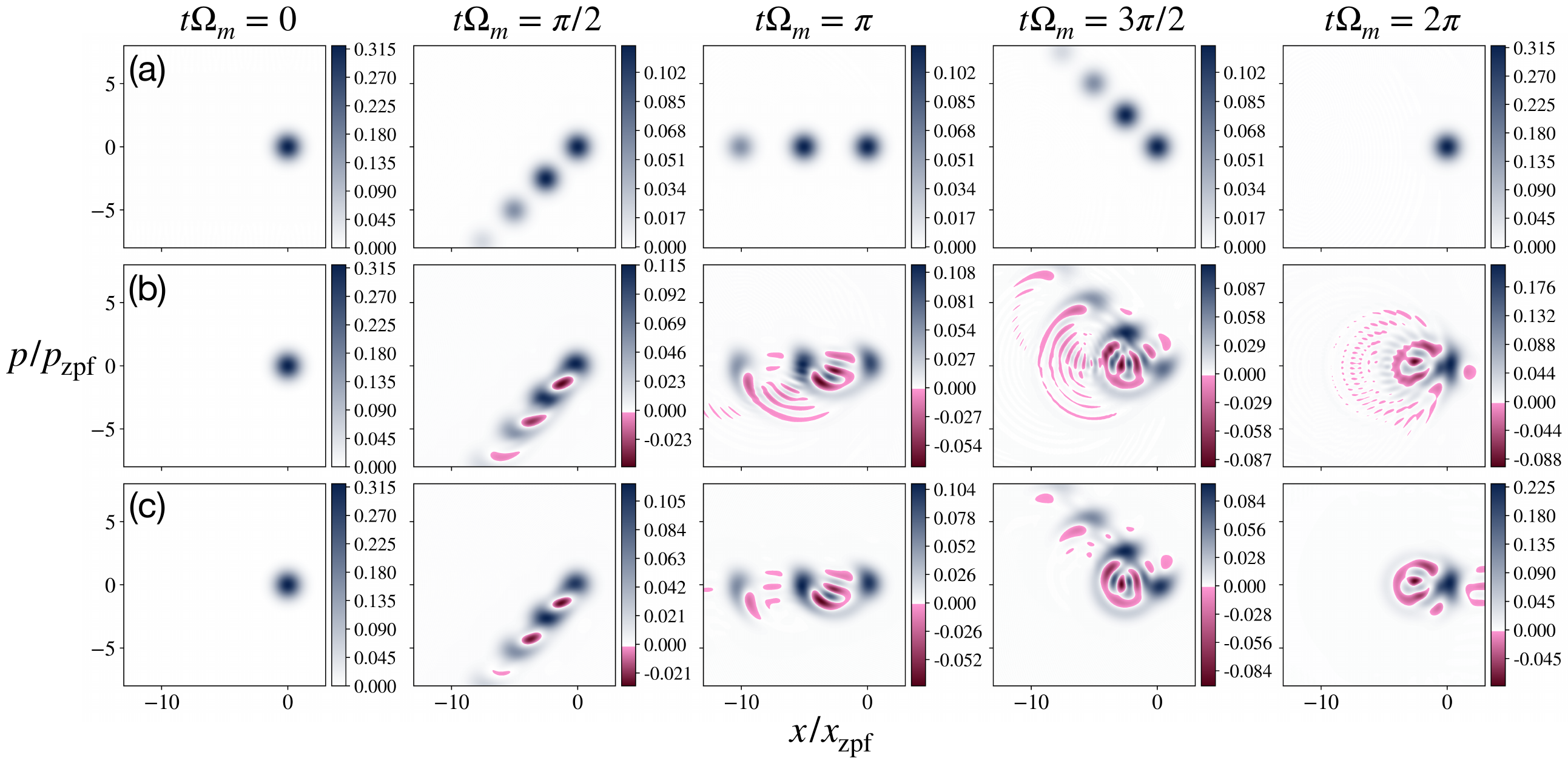} %
\caption{Evolution of the mechanical oscillator Wigner function over one mechanical period. The undriven (a) and driven ($\varepsilon/\Omega_m = 0.3$) (b) cases correspond to Figs.~2(c) and~2(d) in the main text, here with the individual colorbars given. The final row (c) is the Wigner function calculated for the analytical weak drive result, given by Eq.~(\ref{eq:psi_lab}). The initial state is chosen to be $\vert \psi(t=0)\rangle = \vert\alpha = 1\rangle_a\otimes\vert\beta=0\rangle_b$, coupling constant $g_0/\Omega_m = 1.8$ and laser detuning $\Delta = 0$. Note that in the plots presented $x_\mathrm{zpf}$ and $p_\mathrm{zpf}$ are scaled by a factor $\sqrt{2}$.}
\label{fig:Wigner_ev_an} 
\end{figure}

\begin{figure}
\centering
\includegraphics[width = 0.9\textwidth]{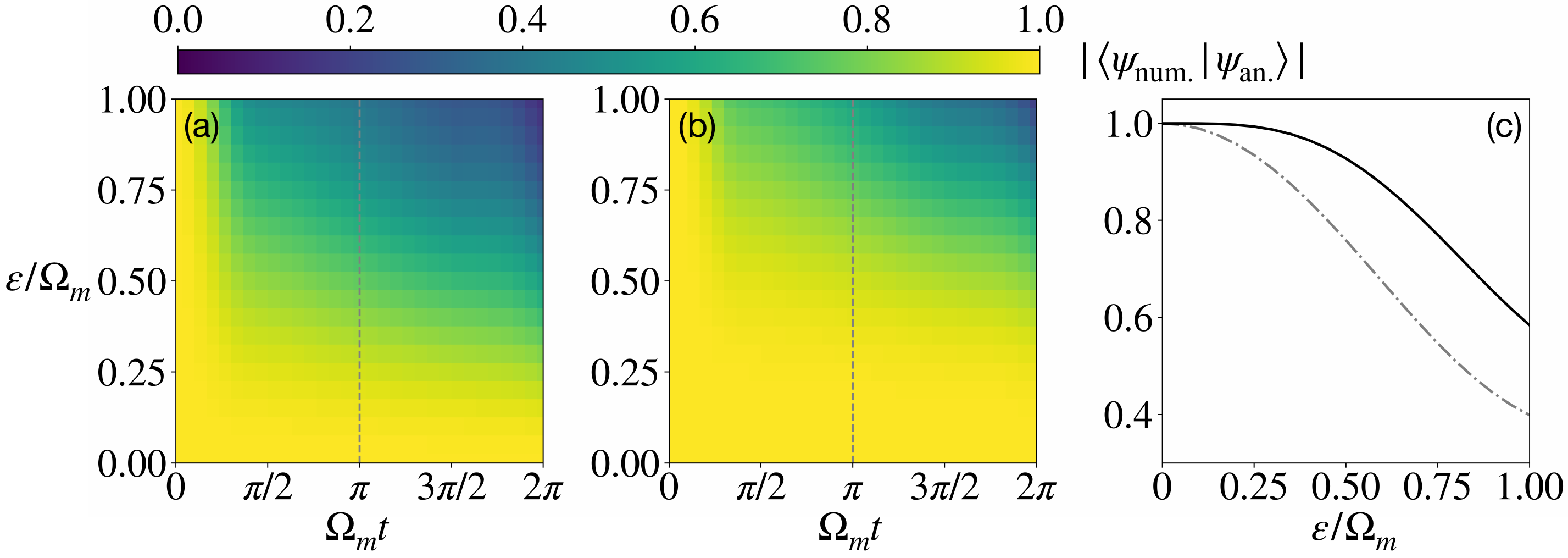} %
\caption{Overlap between the exact numerical solution for the driven problem with (a) the undriven solution and (b) the first order perturbative solution Eq.~(\ref{eq:psi_lab_explicit}), as a function of drive strength and time. (c) A cut at $t=\pi/\Omega_m$ indicating that the perturbative result (solid line) matches the numerical result better than the undriven result (dot-dashed line). The initial state is chosen to be $\vert \psi(t=0)\rangle = \vert\alpha = 1\rangle_a\otimes\vert\beta=0\rangle_b$, with drive strength $\varepsilon/\Omega_m=0.3$ and detuning $\Delta = 0$ and the coupling constant $g_0/\Omega_m = 1.8$.}
\label{fig:overlap_num_an} 
\end{figure}

\begin{figure}
\centering
\includegraphics[width = 0.9\textwidth]{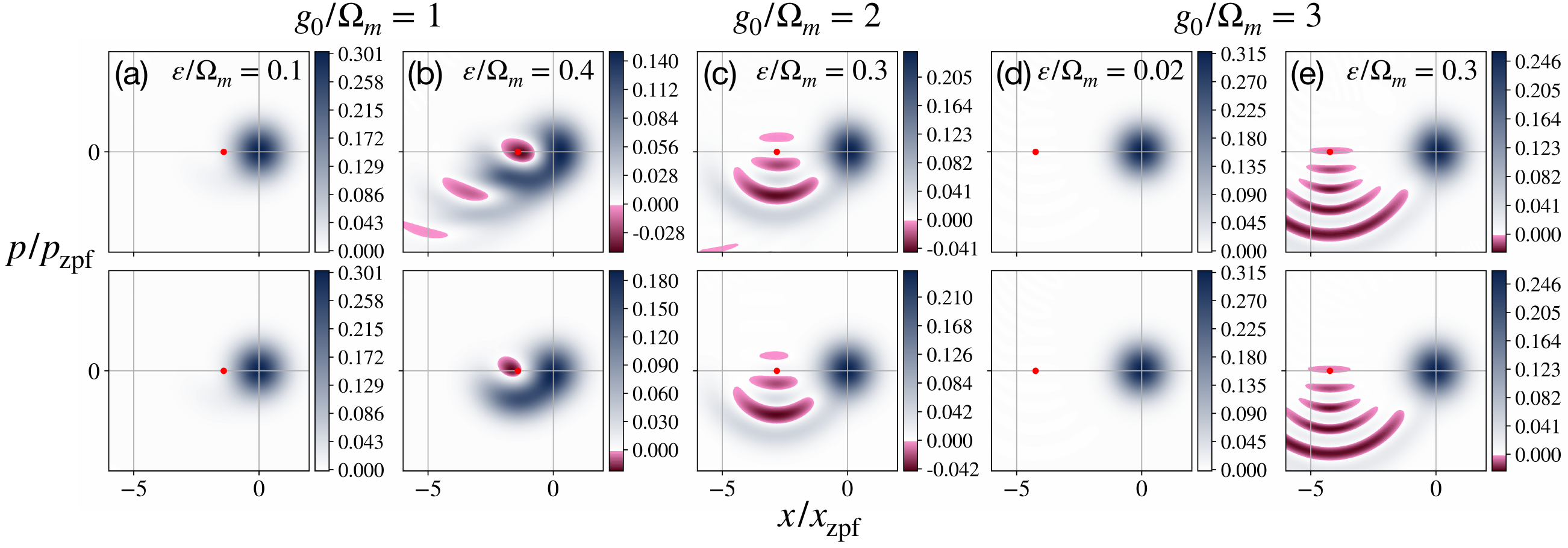} %
\caption{Mechanical Wigner function at $t=\pi/\Omega_m$ after driving for different values of OM coupling $g_0$ and drive power $\varepsilon$, for the initial state $\vert \psi(t=0)\rangle = \vert\alpha = 1\rangle_a\otimes\vert\beta=0\rangle_b$. For $g_0/\Omega_m = 1$ and $g_0/\Omega_m = 3$ we include examples of drive strengths below (a, d) and above (b, e) the threshold to have negativity at the point $\xi = -\tilde{g}_0$ (red dot), as predicted by Eq.~(\ref{eq:W_g0_approx}). In (c) we have  $g_0/\Omega_m = 2$ (even), so in line with Eq.~(\ref{eq:W_g0_approx}) there is no negativity at $\xi = -g_0/\Omega_m$. The top row is the numerical result and the bottom row is the analytical result $W = \mathcal{N}^2 \left[W_0 + \varepsilon^2 W_1 \right]$ with $W_0$ the Wigner function of the vacuum and the correction term $W_1$ being given in Eq.~(\ref{eq:W_1}).}
\label{fig:Wigner_pi_vac} 
\end{figure}

We include here in Figs.~\ref{fig:Wigner_ev_an}(a,b) the same results as presented in Figs.~2(c,d) of the main text, for the evolution of the mechanical Wigner function over one period without and with the laser drive. 
We note that in all Wigner function plots the quantities $x_\mathrm{zpf}$ and $p_\mathrm{zpf}$ are scaled by a factor of $\sqrt{2}$ compared to their standard definition.
Contrary to the main text, here we present the absolute value of the Wigner function for each time, with individual colorbars.
In the additional row Fig.~\ref{fig:Wigner_ev_an}(c) we also include for comparison the Wigner function computed for the state given by our analytical theory Eq.~(\ref{eq:psi_lab_explicit}). 
In both Figs.~\ref{fig:Wigner_ev_an} and~\ref{fig:Wigner_pi_vac} we see that the agreement between numerics and theory is best around the mechanical coherent state components corresponding to the cavity photon states $n=0$ and $n=1$. 
This may be understood from the fact that the first order weak drive expansion only includes events corresponding to adding/subtracting one photon. 
To test quantitatively the conditions of validity of the perturbative result for the wavefunction, we compare to the exact numerical result. 
We compute the overlap $|\langle \psi_\mathrm{num.}|\psi_\mathrm{an.}\rangle|$ as a function of the drive strength and driving time in Fig.~\ref{fig:overlap_num_an}. 

In the main text we present the analytical expression for the Wigner function for the initial state $\vert \psi(t=0) \rangle = \vert \alpha = 0 \rangle_a \otimes \vert \beta =0 \rangle_b$, which reads $W = \mathcal{N}^2 \left[W_0 + \tilde{\varepsilon}^2 W_1 \right]$ with $W_0(\xi) = (2/\pi)e^{-2|\xi|^2}$ where $\mathrm{Re}\xi = x/x_\mathrm{zpf}$ and $\mathrm{Im}\xi = p/p_\mathrm{zpf}$. 
The correction term is:
\begin{equation}
W_1(\xi) = \frac{2}{\pi} \sum_{k=0}^\infty (-1)^k \left| \sum_{m=0}^\infty f_m(t) \langle \xi, k | -\tilde{g}_0, m \rangle\right|^2
\label{eq:W_1}
\end{equation}
where $\vert \xi, k\rangle = \hat{D}(\xi)\vert k \rangle$ is the displaced Fock state.
We defined $f_m(t) = (\tilde{g}_0^m/\sqrt{m!})e^{-\tilde{g}_0^2/2}\left[1-e^{-iE_{1m}t}\right]/\tilde{E}_{1m}$, where $E_{1m}$ is the energy level of the first excited cavity state in the Lang-Firsov frame, $E_{nm} = -\Delta n - \mathcal{K}n^2 + \Omega_m m$, and $\tilde{E}_{nm} = E_{nm}/\Omega_m$.
The overlap of displaced Fock states may be written exactly as \cite{deoliveira1990, wunsche1991}:
\begin{align}
\langle \xi, k | \zeta, m \rangle =&
\left\{\begin{matrix}
\langle \xi|\zeta \rangle \sqrt{\frac{m!}{k!}}(\zeta-\xi)^{k-m} L_m^{k-m}(|\zeta-\xi|^2), & k>m, \\ 
 \langle \xi|\zeta \rangle \sqrt{\frac{k!}{m!}}(\xi^*-\zeta^*)^{m-k} L_k^{m-k}(|\zeta-\xi|^2), & k<m,
\end{matrix}\right.
\end{align}
where $L_m^n$ are the generalized Laguerre polynomials. 
We focus on the point in phase space $\xi=-\tilde{g}_0$ at $t=\pi/\Omega_m$, where the expression simplifies to
\begin{align}
W_1(-\tilde{g}_0) &= \frac{2}{\pi} \sum_{k=0}^\infty (-1)^k \left| f_k(t=\frac{\pi}{\Omega_m})\right|^2 \nonumber \\ 
&= 2\pi\sum_{k=0}^\infty (-1)^k P_k(\tilde{g}_0^2) \left[ \frac{\sin\left(\pi \tilde{E}_{1k}/2\right)}{\left(\pi \tilde{E}_{1k}/2\right)}\right]^2,
\label{eq:W_g0}
\end{align}
where $P_k(\lambda) = \lambda^k e^{-\lambda}/\sqrt{\lambda!}$ is a Poissonian distribution and $\tilde{E}_{1k} = E_{1k}/\Omega_m$.
The Poissonian distribution is maximal at its mean value, $k=\tilde{g}_0^2$. 
The term in square brackets is the $\mathrm{Sinc}$ function, which is maximal when its argument vanishes, at: $k=\tilde{\Delta} + \tilde{g}_0^2$. 
For $\Delta=0$ the peaks of the two functions overlap, leading to $W_1$ giving a significant contribution.
For the particular point $\xi=-\tilde{g}_0$ this contribution may then be positive or negative, and is determined by the parity of the integer closest to $\tilde{g}_0^2$. 
For other points in the $\xi$-plane this condition will be different, but already this indicates the presence of negative areas, and hence explains the origin of the peak in the nonclassical ratio at $\Delta=0$ observed in Fig.~3(a) of the main text. 
At $\xi = -\tilde{g}_0$ for $\Delta=0$, we approximate for the total Wigner function (supposing $\tilde{g}_0^2$ is an integer):
\begin{equation}
W(-\tilde{g}_0) \approx \mathcal{N} \frac{2}{\pi} \left[ e^{-2\tilde{g}_0^2} + \tilde{\varepsilon}^2 \pi^2 (-1)^{\tilde{g}_0^2} \frac{\left(\tilde{g}_0^2\right)^{\tilde{g}_0^2}}{\tilde{g}_0^2!}e^{-\tilde{g}_0^2} \right].
\label{eq:W_g0_approx}
\end{equation}
Taking, for example, $\tilde{g}_0 = 1$, and asking that the total Wigner function be negative, $W(-\tilde{g}_0)<0$, gives the condition on the drive $\tilde{\varepsilon} \gtrsim 0.2$. 
The expression above and argument regarding negativity is in agreement with the full numerical simulation presented in Fig.~\ref{fig:Wigner_pi_vac}.

\begin{figure}
\centering
\includegraphics[width = 0.9\textwidth]{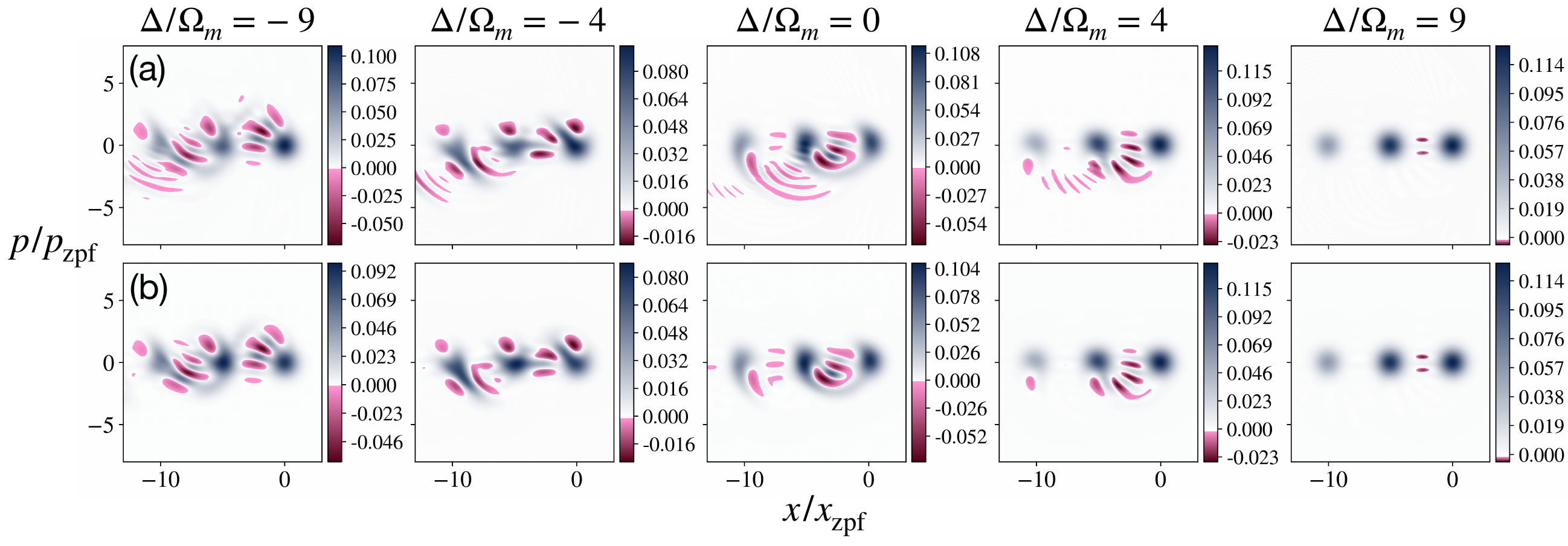} %
\caption{Mechanical Wigner function at $t=\pi/\Omega_m$ for different values of laser detuning. We show the comparison between numerics (a) and our perturbative analytics (b), showing good agreement. The initial state is chosen to be $\vert \psi(t=0)\rangle = \vert\alpha = 1\rangle_a\otimes\vert\beta=0\rangle_b$, drive strength $\varepsilon/\Omega_m = 0.3$ and the coupling constant $g_0/\Omega_m = 1.8$}
\label{fig:Wigner_pi_delta} 
\end{figure}

In Fig.~3(b) of the main text we observe some structure in the parameter plane for the nonclassical ratio $\eta$ for the initial state $\vert \psi(t=0) \rangle = \vert \alpha =1 \rangle_a \otimes \vert \beta =0 \rangle_b$.
We include in Fig.~\ref{fig:Wigner_pi_delta} some plots of the mechanical Wigner function at $t=\pi/\Omega_m$ for fixed drive and coupling strength and different values of the detuning.
This indicates the near vanishing of the Wigner negativity for positive detuning. 

\section{Dissipation}
\setcounter{figure}{0} 
To account for dissipation we solve the optomechanical master equation given by Eq.~(10) of the main text for the density matrix $\rho(t)$. 
To complement the plots given in Fig.~4 of the main text indicating the effect of finite dissipation on the mechanical nonclassicality produced via our driving method, we include here in Fig.~\ref{fig:Wigner_ev_diss} examples of the evolution of the Wigner function over one period of driving for increasing cavity dissipation rate. 
\begin{figure}
\centering
\includegraphics[width = 0.9\textwidth]{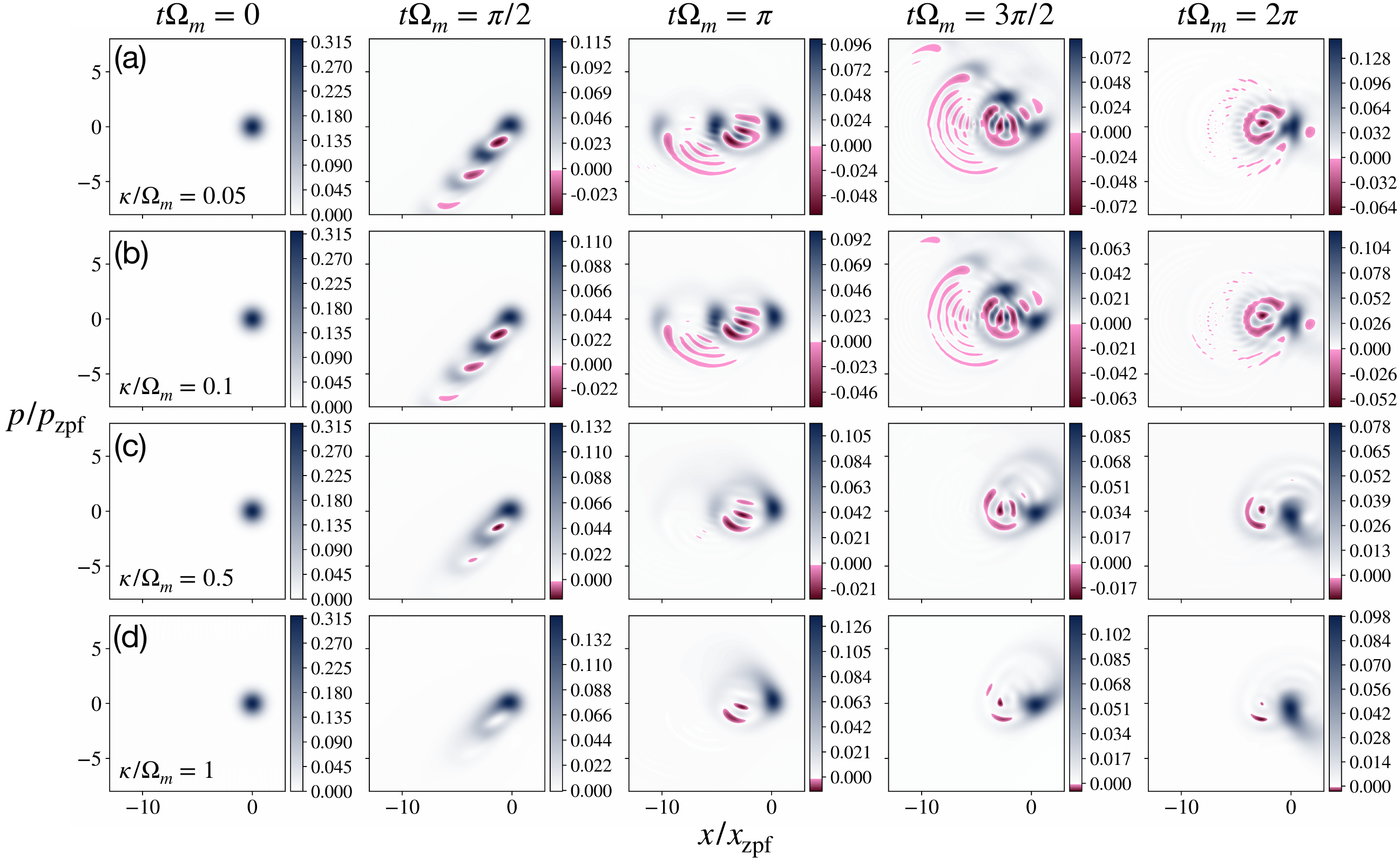} %
\caption{Evolution of the mechanical oscillator Wigner function over one mechanical period for increasing cavity dissipation rate, $\kappa$. The initial state is $\vert \psi(t=0)\rangle = \vert\alpha = 1\rangle_a\otimes\vert\beta=0\rangle_b$ and other system parameters are $\varepsilon/\Omega_m = 0.3$, $g_0/\Omega_m = 1.8$ and $\Delta = 0$, $\Gamma_m/\Omega_m = 10^{-4}$, $\bar{n}_\mathrm{th} = 10$.}
\label{fig:Wigner_ev_diss} 
\end{figure}

\begin{figure}
\centering
\includegraphics[width = 0.5\textwidth]{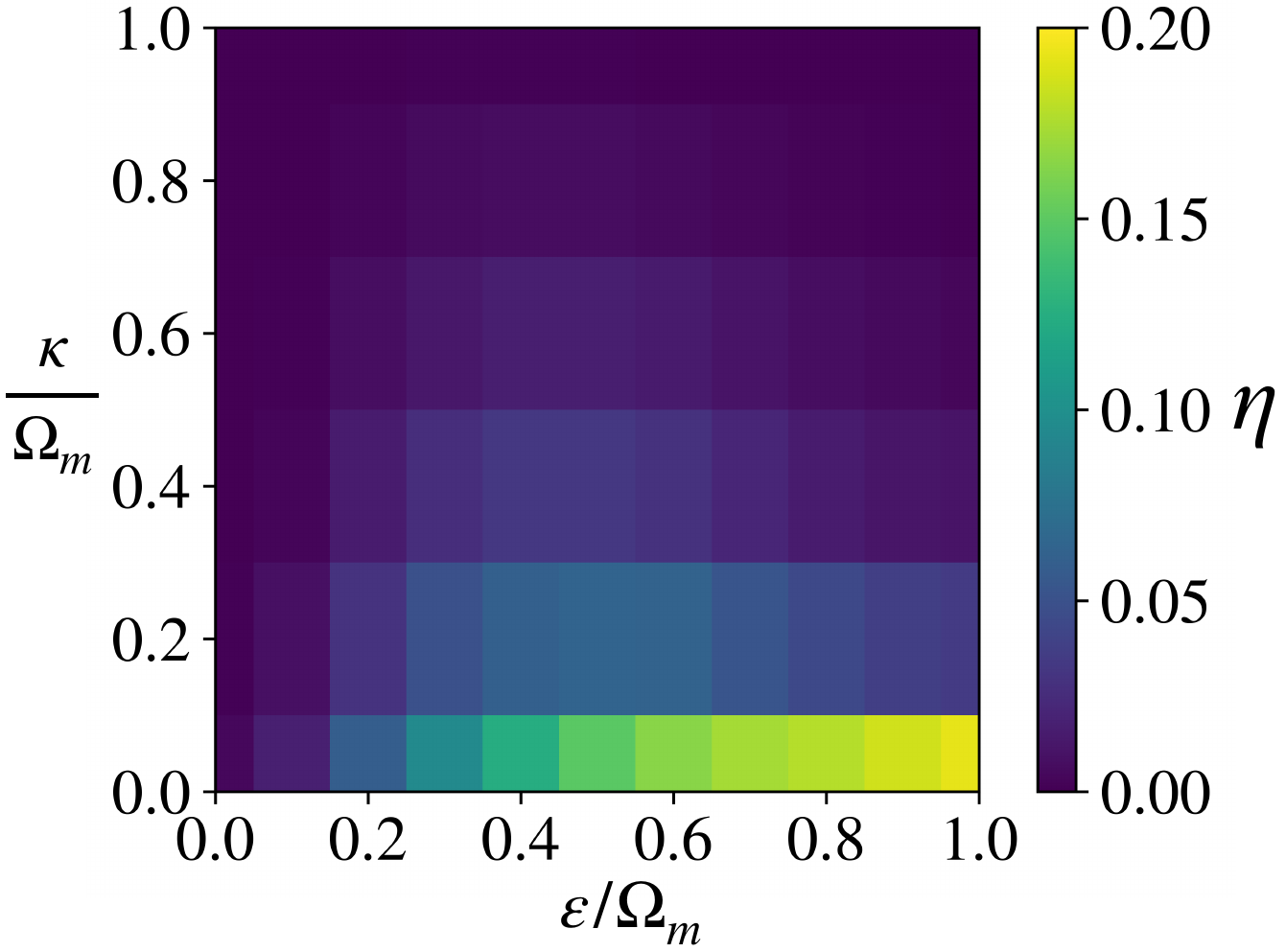} %
\caption{The nonclassical ratio $\eta$ of the mechanical state after driving for half a mechanical period, as a function of the drive strength $\varepsilon$ and the cavity dissipation rate $\kappa$. This figure shows the original data corresponding to the interpolation presented in Fig.~4(a) of the main text. The initial state is $\vert \psi(t=0)\rangle = \vert\alpha = 1\rangle_a\otimes\vert\beta=0\rangle_b$ and other system parameters are $g_0/\Omega_m = 1$ and $\Delta = 0$, $\Gamma_m/\Omega_m = 10^{-4}$, $\bar{n}_\mathrm{th} = 0$.}
\label{fig:NC_diss_raw} 
\end{figure}

Fig.~4(a) of the main text is created by interpolation of the results of a limited number of numerical simulations. 
We include here the original data in Fig.~\ref{fig:NC_diss_raw}.


\bibliography{SM}